\documentclass[11pt,a4paper]{article}
\usepackage{jheppub}
\usepackage[utf8]{inputenc} 
\usepackage{multirow}    
\usepackage{xcolor}
\usepackage{graphicx}
\usepackage{amsmath}
\usepackage{amssymb}
\usepackage{braket} 
\usepackage{hyperref}
\usepackage{frontespizio}
\usepackage{comment}  
\usepackage{epsfig}
\usepackage{float}
\usepackage{bbold}
\usepackage[title]{appendix}
\usepackage{multirow} 
\usepackage{tikz}
\usetikzlibrary{decorations.pathmorphing}
\usepackage{varwidth}
\usepackage{tikz}
\usetikzlibrary{backgrounds,patterns,calc}
\usepackage{xparse}  
\usepackage{subfigure}  
\usepackage{bbm}   
\usepackage{mathabx}

\usepackage{jheppub} 

\title{Particle Physics: a crash course for Mathematicians}

\author[]{Veronica Pasquarella}
\affiliation{Department of Applied Mathematics and Theoretical Physics (DAMTP)}
\affiliation{University of Cambridge, \\ Wilberforce Road, CB3 0WA, Cambridge, UK}

\emailAdd{vp360@damtp.cam.ac.uk}

\abstract{This introductory work combines bottom-up and top-down approaches towards understanding the underlying categorical structure of possible unifying theories descending from string theory. Guided by well-established developments in the realm of categorical algebraic geometry, we explain why abelianisation could potentially lead to furthering the understanding of how to embed Beyond the Standard Model scenarios in supersymmetric setups.}

\makeatletter
\gdef\@fpheader{}
\makeatother
\begin{document} 
\maketitle

\section{Introduction}

\medskip  

\medskip 

At first sight, the title chosen for this article might appear provocative, and the length of the paper is certainly not fair towards the rich and deep advancements made by the phenomenological community so far\footnote{For a detailed overview of the state of the art and major developments in the field, we refer the interested reader to the rich literature at our disposal, including \cite{Nilles:1995ci,Csaki:1996ks,Slavich:2020zjv,LHCReinterpretationForum:2020xtr,Allanach:2021bbd,Banks:2020gpu,Allanach:2023bgg,Hammou:2023heg,Kassabov:2023hbm,Iranipour:2022iak,Nath:2010zj}, and \cite{Allanach:2005pv,Cicoli:2021dhg,Krippendorf:2010hj,AbdusSalam:2009qd,AbdusSalam:2007pm,Conlon:2007xv,Cremades:2007ig} for slightly more theoretically-oriented perspectives in this regard, especially in relation with string theory, \cite{gsw1,gsw2,jp1,jp2,Green:1982ct,Schwarz:1982jn,Green:1984sg,Gross:1984dd,Polchinski:1994mb}.}. However, we stress that our aim is that of introducing important open questions in the realm of Particle Physics in a mathematical language approachable by categorical algebraic geometers.

Actually, this work is really meant to be the first of a series of papers by the same author continuing from past developments, bridging the gap between Pure Mathematics and Theoretical Particle Physics achievements. The author's engagement with experts from both lines of research is what motivated this project in the first instance. In upcoming work, \cite{VP}, we will be developing this description in terms of algebraic and Hilbert series calculation in greater detail. For the moment, we anticipate that the purpose of this introductory article is mainly fourfold:  

\begin{itemize}  

\item   To illustrate (some) key fundamental problems in Theoretical Particle Physics necessitating further mathematical understanding.   

\item   Opening the scene to well-established techniques in representation theory and categorical algebraic geometry to help in this regard.    

\item   For those who are not experts in the field, serve as an oversimplified explanation of the powerful formalism put forward by Braverman, Finkelberg, and Nakajima (BFN), \cite{Braverman:2017ofm,Braverman:2016wma}. In doing so we will explicitly show the relation in between their work, that of Freed, Moore, and Teleman (FMT), \cite{Freed:2022qnc}, and Teleman's, \cite{Teleman:2014jaa}, which, to the best of our knowledge, has only previously been referred to in \cite{Pasquarella:2023exd,Pasquarella:2023}. 

\item Motivating Phenomenologists to engage more with Pure Mathematicians, fostering mutual inspiration for common aims.

\end{itemize}

The main objective of this paper is that of explaining the higher categorical structures that are needed for describing the invariants associated to specific supersymmetric quiver gauge theories, with a particular focus on dualities and their mutual relations in terms of higher-categories.  We believe that deepening its understanding and generalisations thereof, could lead to uncharted corners of mathematics that could be used for embedding of the Standard Model (SM) of Particle Physics, \cite{Weinberg:2004kv} in possible supersymmetric unifying theories, \cite{Quevedo:2010ui}.

The present work is structured as follows: 

\begin{enumerate}

\item Section \ref{sec:2}  explains the motivations and outline of the present and forthcoming papers \cite{VP}. In doing so, we briefly introduce scattering amplitudes as crucial tools within the context of particle physics, explaining their connection with Hilbert series. We then briefly explain some of the key open questions in Theoretical Particle Physics necessitating further mathematical understanding, specifically the hierarchy problem, and the lightness of the Higgs mass (and its relation to Supersymmetry (SUSY), \cite{Quevedo:2010ui}).

\item     Section \ref{sec:4}   is mostly a revision of \cite{Freed:2022qnc,Freed:2012bs}, and we use it at this stage for multiple reasons. First of all because it allows to introduce gauge theories and Lagrangians in a mathematical language, which in turn is suitable to make connection with the higher-categorical Symmetry Topological Field Theory (SymTFT) prescription put forward by Freed, Moore, and Teleman, (FMT), \cite{Freed:2022qnc}, in the context of relative quantum field theories (QFTs).  

\item     The natural way of describing gauge theories is by means of quivers. The Standard Model itself admits such a representation. It is also the case, though, that quivers are crucial for describing certain algebraic varieties associated with highly supersymmetric setups. Section \ref{sec:5} is therefore devoted to explaining in a simplified way the well-established BFN construction leading to ring homologies. The theories in question are 3D ${\cal N}=4$ supersymmetric quiver gauge theories. This section highlights the importance that categorical algebraic geometry plays in identifying ring homologies accounting for the complete spectrum of a given theory, therefore giving a complementary insight with respect to the Hilbert series alone. In so doing, we will highlight the importance of \emph{abelianisation}, \cite{Dimofte:2018abu,Bullimore:2015lsa}.  In doing so we will explicitly show the relation in between their work, that of Freed, Moore, and Teleman (FMT), \cite{Freed:2022qnc}, and Teleman's, \cite{Teleman:2014jaa}, which, to the best of our knowledge, has only previously been mentioned in \cite{Pasquarella:2023exd,Pasquarella:2023}. 

\item From the discussion of the previous section, the concluding section, \ref{sec:6}, paves the way to forthcoming work \cite{VP}, where a more detailed analysis with regard to the algebraic structure and Hilbert series of (Beyond) the Standard Model scenarios will be carried out.  
Additional complementary directions addressed by the same author will be developing ring homologies for Moore-Tachikawa varieties, \cite{Moore:2011ee}, beyond categorical duality\footnote{Thereby following up from the analysis carried out in \cite{Pasquarella:2023ntw}.}, and understanding their implications for Koszul duality, \cite{Webster:2016rhh}. We plan to report of any advancements in this regard in the near future, \cite{VP}.

\end{enumerate}

\section{Roadmap}    \label{sec:2}

\begin{figure}[ht!]  
\begin{center}  
\includegraphics[scale=1]{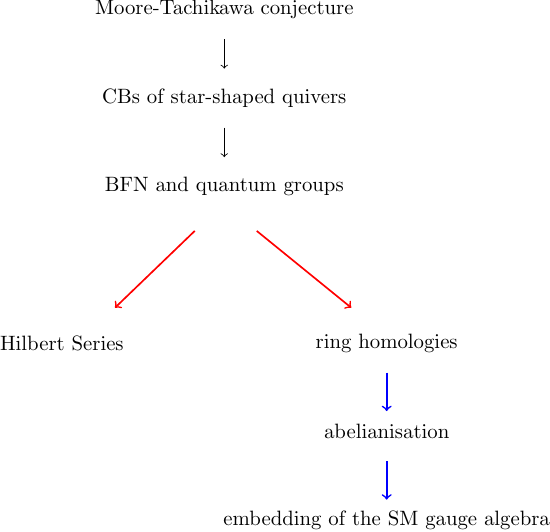}  
\label{fig:roadmap}  
\caption{\small Roadmap for the present and upcoming work.}  
\end{center}  
\end{figure}

This first section is devoted to outlining the path we will be adopting for addressing the issues mentioned in the Introduction. The diagram of figure \ref{fig:roadmap} exemplifies this, and we will start addressing the topics outlined there starting from section \ref{sec:5}, continuing in an upcoming work \cite{VP}. For those who are already familiar with the terminology adopted in figure \ref{fig:roadmap}, the first three sections of this paper are purely revision. However, they might still find some interesting connections, which, to the best of our knowledge, haven't been spelled out in the literature with the same emphasis and perspective.

Essentially, sections \ref{sec:2}, \ref{sec:4}, and \ref{sec:5} can be thought of as an extremely concise summary of the basic knowledge gathered by a mathematical physicist in the attempt to explain a possible roadmap towards deepening the understanding of some major open questions in Theoretical Particle Physics from a Mathematician's point of view. 

As an advanced notice, we advice the reader not to be misguided by the oversimplified explanation of the first few pages of this work: technicalities will soon arise.

In this preliminary section, we introduce the unfamiliar reader to: 

\begin{enumerate}  

\item The basic building blocks of QFT\footnote{We strongly suggest the interested reader to refer to the work of pioneers in the field for a complete treatment of QFT, specifically \cite{W1,W2,W3}.}.  

\item Some key theoretical tools adopted by Phenomenologists for analysing them.   

\end{enumerate}

\subsection{Fields and Interactions}

Quantum fields are mutually-interacting entities exhibit wave-particle duality. QFT is arguably the most successful scientific theory so far, with the Standard Model (SM) of Particle Physics being a prototypical example. In some cases, QFT predicts to great accuracy observational outcomes, and has also been used in the context of pure mathematics. 

With the advent of the quantum revolution, fields were promoted to being fundamental constituents, whereas all fundamental particles were understood to arise as a quantum mechanical (QM) effect. This extends to all particles known so far, namely quarks, leptons, gluons, W- and Z-bosons, and the Higgs boson. 

The SM is an example of a QFT containing all the particles known up to now, with their associated three force fields\footnote{The matter fields comprised in it come in 3 families, plus the Higgs boson.} Extensions of the SM, referred to as Beyond the Standard Model (BSM) setups, are currently an active area of research, providing extensions of the SM by adding interactions and/or particles. 

A given QFT is defined in terms of a Lagrangian density, which, upon integrated over spacetime, defines the action, $S$, which is, essentially, a functional of the fields. The Variational Principle constraints the action in such a way that its extremisation leads to the equations of motion of the respective fields. Importantly, the action is needed for defining the partition function, i.e. the expectation value of the identity, as well as any other expectation value of operators\footnote{The latter are also known as correlation functions. } accounted for by a given effective field theory (EFT).

\subsection{Amplitudes and Hilbert series} 

\subsection*{Amplitudes}  

Scattering amplitudes are among the most important tools in Particle Physics. They are the building blocks for defining transition amplitudes of a certain set of states to another. In the context of collider physics, they are an essential theoretical background tool for predicting and analysing particle collisions and decays. Importantly, if new physics were to be discovered, it must be signalled by a mismatch in between experimental outcomes and scattering calculations, that could either feature as the detection of new particles, and/or the identification of a new interaction (and therefore a new fundamental force). 

\begin{figure}[ht!]  
\begin{center}  
\includegraphics[scale=0.8]{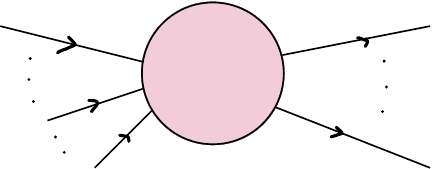}  
\label{fig:scatt}  
\caption{\small A simplified depiction of a scattering process. The $S$-matrix defines the amplitude from the incoming states (black lines on the left) to the outgoing states (black lines on the right). The shaded circle in the middle encodes the information regarding the interactions and operator content characterising a certain EFT.}  
\end{center}  
\end{figure}

In a nutshell, if our theoretical knowledge of particle physics were complete, we should be able to explain all experimental outcomes, recasting them to the Lagrangian theories at hand. However, it so happens that there are some crucial unanswered questions, that clearly signal the need for a deeper mathematical understanding of the models developed so far. Those that motivated our work in the first place are the following:  

\begin{enumerate}  

\item The mass hierarchy problem.   

\item The lightness of the Higgs mass. 

\end{enumerate} 

To help the unfamiliar reader, we will come back to explaining these issues in section \ref{sec:3}.

\subsection*{Hilbert series and effective operators}  

\begin{equation}  
\text{HS}[\phi]\ =\ \prod_{_{i=1}}^{^{n}}\ \int_{_{{\cal G}_{_{i}}}}d\mu_{_i}\ \text{PE}\ [\varphi, R]  
\label{eq:HS}  
\end{equation}   
where ${\cal G}_{_i}$ denotes a given gauge group, $\varphi$ is a \emph{spurion} variable\footnote{A suprion could either be a scalar, a fermion, or a gauge field.}, $d\mu_{_i}$ denotes the Haar measure, and PE$[{\varphi, R}]$ the \emph{plethystic exponential}, which is defined as follows

\begin{equation}  
\text{PE}\ [\phi, R]  =\exp\left[\sum_{r=1}^{\infty}\frac{\phi^{^r}\chi_{_R}(z_{_j}^{^r})}{r}\right]\ \ \ \ \ \ \ \ ,\ \ \ \ \ \ \ \ \text{PE}\ [\psi, R]  =\exp\left[\sum_{r=1}^{\infty}(-1)^{^{r+1}}\frac{\psi^{^r}\chi_{_R}(z_{_j}^{^r})}{r}\right]
\end{equation} 
for bosons and fermions, respectively. We will be entering the details of the calculation of \eqref{eq:HS} in an upcoming work, \cite{VP}. For the purpose of this introductory work, it is enough to know that HS is constructed in terms of the degrees of freedom (namely the particle content) and their transformation properties under the symmetries of a given theory. This is enough to say that, in order to build the Hilbert series, we need to know what the complete set of operators of a given theory is. Completeness of the spectrum of operators is a major sought-after objective in particle physics, since it paves the way towards identifying possible extensions of the fundamental laws of nature. Given the relation in between scattering amplitudes and HS, a mismatch with predictions for a given theory signal the need to extend the effective field theory of reference. If the theory in question is taken to be the Standard Model (SM) of Particle Physics, its resulting extensions are referred to as Beyond the Standard Model (BSM) effective field theories, \cite{Graf:2022rco,Bento:2023owf,Anisha:2019nzx}.

\medskip    

\medskip

\subsection{(Some) Open Questions in Quantum Field Theory}  \label{sec:3}  

Albeit having been successfully verified experimentally, the SM of Particle Physicsis clearly incomplete. The Higgs boson, \cite{Higgs:1964ia,Higgs:1966ev}, responsible for the mass of fundamental particles, was the last piece of a jigsaw to be found, yet the first of a more complicated one. The Higgs boson is a scalar responsible for the mass gain of fundamental particles, via the so-called Higgs mechanism\footnote{We refer the interested reader to the original work by Professor Peter Higgs for a detailed explanation of the process, \cite{Higgs:1964ia,Higgs:1966ev}.}. Undoubtedly one of the major theoretical physics achievements of the past century, the model is clearly incomplete. The two major issues we indicate are the mass hierarchy problem and the lightness of the Higgs boson mass. Each one of them is the object of active investigation by world-leading experts in the field, and by no means we are attempting to provide a full explanation of either of them in our treatment. Nevertheless, we consider them as part of the motivations leading to the need for furthering the understanding of the underlying mathematical structure of QFT.

\subsection*{The Mass Hierarchy Problem}  

Essentially, this is a statement regarding the unexplainable discrepancy between the energy scales at which the fundamental interactions decouple. A successful unifying theory should be able to explain why this symmetry breaking pattern. 

\subsection*{Lightness of the Higgs Boson and Supersymmetry}

Another open question highlighting the fact that the SM is still incomplete is that the Higgs mass is too light. Given that it is the only spinless particle in the SM, theorems imply it should be extremely massive. One of the possible reasons explaining the very light mass of the Higgs was thought to be Supersymmetry (SUSY), proposing that, in order to compensate the very light mass observed, new particles (referred to as superpartners) should be detected at the Large Hadron Collider (LHC). Up to now, though, these superpartners haven't been observed by collider experiments at the energies predicted by the theory. This inevitably leads to the question, to what extent is SUSY able to provide a consistent theory of everything, given that String Theory so strongly relies upon it?  

We believe that SUSY and String Theory are very promising candidates for a unifying theory of everything, but, in order to overcome the apparent shortcomings listed above, we need to address the deep question highlighted in the introduction, namely: "What is (the mathematical structure of) a QFT?". The present work is meant to provide further insights in addressing this question, by making use of mathematical advancements in the field of categorical algebraic geometry. Our findings shed light on a new perspective that is proposed to be a promising alternative top-down construction of QFTs. The ultimate aim would be that of identifying the mathematical structure underlying the SM, and, with it, its embedding in the most suitable string theory vacuum possible, to which we plan to report in upcoming work \cite{VP}.

\section{Quantum Field Theory from Category Theory}   \label{sec:4}

We now turn to explaining how QFT can be mathematically described in terms of Category Theory. In doing so, we will also be introducing the notion of moduli space, as denoting the minima of the theory in question, namely the least energy field configurations. Albeit the sample theories presented in this section are quite trivial, they are sufficient for building intuition for what lies ahead in our treatment. 

To begin with, we start off by listing some of the ingredients that are most relevant to our analysis\footnote{As we shall see, more richness will be added in due course in our treatment. The aim of the table is simply that of helping the reader in gaining intuition for the setup and the motivations behind the procedure adopted.}, briefly outlined in the table below.

\begin{equation}
\begin{aligned}
&\text{\underline{Particle Physics}} \ \ \ \color{white}spacespacespacespace\color{black} \ \ \text{\underline{Category Theory}} \\
&\\
&\text{Gauge Theory}\ \ \ \color{white}spacespacespacespaces\color{black} \ \ \text{Algebras}\\
&\text{\color{red}Operators}\ \ \ \color{white}spacespacespacespacespace\color{black} \ \ \text{Homologies}\\
&\text{Quivers}\ \ \ \color{white}spacespacespacespacespacesp\color{black} \ \ \text{Quivers/Algebraic varieties}\\
&\text{Hilbert Series}\ \ \ \color{white}spacespacespacespacece\color{black} \ \ \text{Hilbert Series}\\
&\text{\color{red}SUSY}\ \ \ \color{white}spacespacespacespacecespacesp\color{black} \ \ \text{SUSY}\\
\end{aligned}
\nonumber
\end{equation}  

\medskip 

\medskip 

This shows some quantities/entities encountered in the realm of Particle Physics and their Category Theory counterpart. We have highlighted two of them in red. The reason for this will be more clear in section \ref{sec:5}, but the familiar reader will certainly notice that some motivation was already hinted at in section \ref{sec:2}. For the moment, we simply recall what was mentioned in the Introduction, namely that the Hilbert series and ring homologies are key complementary ingredients towards understanding a given QFT. Among the two, though, the latter is the one that is more directly related to understanding the underlying categorical structure of the theory in question, and therefore constitutes the ultimate objective one wishes to obtain. On the other hand, the Hilbert series, is obtained from ring homologies, but hides some key categorical features that are only accessible via the homological perspective.\footnote{Indeed, it is known that different ring homologies could lead to the same Hilbert series, \cite{Cremonesi:2014vla}.}

This intermediate section is mostly an overview of \cite{ Freed:2022qnc,Freed:2012bs,Freed:2022qnc}, and is needed in order to understand the analysis carried out in the concluding part of our work, \ref{sec:5}. This section is structured as follows:  

\begin{enumerate}  

\item At first, we review mathematical gauge theory and the definition of moduli spaces therein, \cite{Freed:2022qnc}. 

\item Then, we recall the notion of relative field theories, \cite{Freed:2012bs}, which can also be adapted to gauge theories. 

\item In conclusion, we explain how this fits with the higher-categorical prescription of Freed, Moore and Teleman, \cite{Freed:2022qnc}, and how gauging can be implemented in terms of condensable homomorphisms in the Symmetry Topological Field Theory categorising the symmetries of a given QFT, \cite{TJF}. 

\end{enumerate}  

For reasons that will become clear by the end of this section, the crucial part of the higher-categorical perspective can be summarised as follows: 

\begin{itemize}  

\item Keeping track of the identity. 

\item Embedding the gauged algebra. 

\end{itemize}  

Most importantly, they both lie within the notion of \emph{abelianisation}. In section \ref{sec:5} we will see explicitly how they arise within the context of supersymmetric quiver varieties, \cite{Dimofte:2018abu,Bullimore:2015lsa}.

\subsection{Principal bundles and connections} 

Formulating gauge theories in Lagrangian formulation that is suitable for a QFT requires some preliminary differential geometry notation, which we will now overview.  

The initial ingredients we start from are a manifold, $M$, and a Lie group, $G$. Then, a \emph{principal G-bundle}, $P\rightarrow M$ is defined as a manifold $P$ on which $G$ acts freely on the right with quotient 

\begin{equation} 
P/G\simeq M,  
\end{equation}  
such that there exist local sections. If $P^{^{\prime}}, P$ are principal $G$-bundles over $M$, then an isomorphism  

\begin{equation}  
\varphi^{^{\prime}}:P^{^{\prime}}\ \longrightarrow\ P,     
\end{equation}   
is a smooth diffeomorphism which commutes with $G$, and induces the identity map on $M$. If $P\equiv P^{^{\prime}}$, such automorphism defines a gauge transformation of $P$.

For every $M$, there is a category of principal $G$-bundles and isomorphisms. 

A \emph{connection} on a principal $G$-bundle $\pi: P\rightarrow M$ is a $G$-invariant distribution in $TP$ which is transverse to the vertical distribution kerd$\pi$. A connection can therefore be expressed as a 1-form $A\in\Omega^{^1}(P;\mathfrak{g})$ whose value at $p$ is the projection

\begin{equation}    
T_{_p}P\ \longrightarrow\ V_{_p}\simeq\mathfrak{g},      
\end{equation}   
with kernel $H_{_p}$.

Connections form a category, arranging in a set of equivalence classes under isomorphisms.

\subsection{Mathematical Gauge theory}  

The prototypical example of a gauge theory is Maxwell's theory. In its quantised formulation, one needs to introduce Dirac's charge quantisation, replacing the group $\mathbb{R}$ by the compact group $\mathbb{Z}/2\mathbb{Z}$. This statement can be generalised by replacing $\mathbb{Z}/2\mathbb{Z}$ with any compact Lie group $G$. 

Hence, we are drawn to the conclusion that the Lagrangian formulation of the field theory in question is simply based on the following data:   

\begin{enumerate}  

\item A compact Lie group with Lie algebra, $\mathfrak{g}$.

\item  A bi-invariant inner product on $\mathfrak{g}$, $<.,.>$.

\end{enumerate}

From this, the total Lagrangian for the gauge field reads

\begin{equation}  
L=-\frac{1}{2}\left<F_{_A}\wedge\star F_{_A}\right>  
\end{equation}
from which the energy density reads   

\begin{equation}  
{\cal E}(A)=\left[\ \sum_{_{\mu<\nu}}\frac{1}{2}\left|F_{_{\mu\nu}}\right|^{^2}\right]\ \left|d^{^{n-1}}x\right|,      
\end{equation}  
whose moduli space is simply 

\begin{equation}  
{\cal M}_{_{vac}}=\text{pt},       
\end{equation} 
where the calculation is understood to be performed on the space of equivalence classes of connections.

\subsection*{The gauged sigma-model}  

This is a very general bosonic field theory, combining pure gauge theory with the $\sigma$-model, both of which are greatly used in QFT. The data specifying the model in question are: 

\begin{enumerate}

\item A Lie group, $G$, with Lie algebra, $\mathfrak{g}$. 

\item A bi-invariant scalar product, $<,>$, on $\mathfrak{g}$.    

\item A Riemannian manifold, $X$, on which $G$ acts by isometries.

\item A potential function   

\begin{equation}    
V:\ X\ \longrightarrow\ \mathbb{R},     
\end{equation}
invariant under $G$.  

\end{enumerate}  

The space of fields, ${\cal F}$, consists of pairs 

\begin{equation} 
{\cal F}\ \overset{def.}{=}\{(A,\phi)\},    
\end{equation}  
with $A$ denoting a collection of principal $G$-bundles $P\rightarrow M$, and $\phi$ a section of the associated bundle over $P\times_{_G}X\rightarrow M$. In most cases, it is convenient to view $\phi$ as an equivariant map 

\begin{equation}    
\phi:\ P\ \longrightarrow\ X.   
\end{equation}

The space of fields can be thought of as a category, with an isomorphism 

\begin{equation}    
\varphi:\ P^{^{\prime}}\ \longrightarrow\ P     
\end{equation}
of principal bundles, inducing an isomorphism of fields  

\begin{equation}     
\left(A^{^{\prime}},\phi^{^{\prime}}\right)\ \longrightarrow\ (A,\phi).  
\end{equation}    

The global symmetry group of the theory is the subgroup of isometries of $X$ commuting with the $G$-action, and preserving the potential, $V$. 

The Lagrangian density, therefore reads  

\begin{equation}    
{\cal L}=-\frac{1}{2}|F_{_A}|^{^2}+\frac{1}{2}|{\cal D}_{_A}\phi|^{^2}-\phi^{^*}V,  
\label{eq:Lagrangiandensity}    
\end{equation}  
where ${\cal D}_{_A}$ denotes the covariant derivative, coupling $\phi$ with $A$. The energy density of the pair $(A,\phi)$, can readily be extracted from \eqref{eq:Lagrangiandensity}, and reads  

\begin{equation}   
{\cal E}(A,\phi)=\left[\ \sum_{_\mu<\nu}\frac{1}{2}|F_{_{\mu\nu}}|^{^2}+\sum_{_{\mu}}\frac{1}{2}\left|\left(\partial_{_A}\right)_{_{\mu}}\phi\right|^{^2}+\phi^{^*}V\ \right]\ \left|d^{^{D-1}}x\right|,     
\label{eq:energyd}   
\end{equation} 
where $D$ denotes the total number of spacetime dimensions. 

Assuming $V$ has a minimum at 0, one can look for soulitions with 0-energy. By direct inspection of \eqref{eq:energyd}, it is possible to deduce that:   

\begin{enumerate}  

\item  The first term implies that $A$ is flat, hence, up to equivalence, it is the trivial connection with zero-curvature.

\item   The second term tells us that $\phi$ has to be a constant for it to be a zero-energy solution.

\item   The third term constraints the constant value of $\phi$ to lie in the $V^{^{-1}}(0)$ set.

\end{enumerate}   

A trivial connection $A$ has a group of automorphisms that are isomorphic to $G$, i.e. the group of global gauge transformations. For a vacuum solution, $\phi$ must be a constant function with values in $V^{^{-1}}(0)$, and, therefore, $V^{^{-1}}(0)$ inherits the same $G$-action from $\phi$. Hence, the moduli space of vacua is a subquotient space of $X$  

\begin{equation}   
\boxed{\ \ \ {\cal M}_{_{vac}}\ \overset{def.}{=}\ V^{^{-1}}(0)\bigg/G\color{white}\bigg]\ \ }.  
\end{equation}

For the case of certain supersymmetric field theories, $X$ is K$\ddot{\text{a}}$hler of hyperk$\ddot{\text{a}}$hler, and $V^{^{-1}}(0)$ is the norm square of an appropriate moment map, in which case ${\cal M}_{_{vac}}$ reduces to a K$\ddot{\text{a}}$hler of hyperk$\ddot{\text{a}}$hler quotient (with the latter actually being the setting where this had first been studied).

\subsection{Relative Field Theories}

Symmetries in field theory are expressed in terms of topological defects acting as operators on state spaces of the theory in question. The idea at the heart of this are the notions of abstract groups and abstract algebras. In their work, \cite{Freed:2022qnc}, Freed, Moore and Teleman propose an abstract symmetry structure in the context of field theory and its concrete realisation.

A field theory is analogous to a linear representation of a Lie group or a module over an algebra. Hence their definition of the boundary theory as a module. The essential content of their definition is that of expressing a QFT, $F$, as a sandwich $\rho\otimes_{\sigma}\tilde F$, with $\rho$ a regular Dirichlet BC (right boundary theory for $\sigma$) and $\tilde F$ the left boundary theory for $\sigma$ (typically not topological). 

Altogether, $(\sigma,\rho)$ constitutes the $n$-dimensional \emph{quiche}, defining the abstract symmetry structure of a given QFT. Defects away from $\tilde F$ are topological and belong to $(\sigma,\rho)$. $(\sigma,\rho)$-defects are analogous to elements of an abstract algebra (connection with GKSW). The novelty of the treatment by Freed, Moore and Teleman is the fact that the sandwich representation develops a calculus of topological defects acting on a QFT (based on fully local TFT). The TFT imposes strong finiteness constraints on the QFT. 

Fixing $N\in\mathbb{Z}^{^{\ge 0}}$, then a quiche is a pair $(\sigma, \rho)$ in which

\begin{equation}   
\sigma:\text{Bord}_{_{N+1}}(F)\rightarrow{\cal C} 
\end{equation} 
is an $N+1$-dimensional TFT and $\rho$ is a right topological $\sigma$-module. The quiche is $N$-dimensional, hence it shares the same dimensionality as the theory on which it acts. Let $F$ be an $N$-dimensional field theory. A $(\sigma,\rho)$-\emph{module structure} on $F$ is a pair $(\tilde F, \theta)$, in which $\tilde F$ is a left $\sigma$-module and $\theta$ is an isomorphism 

\begin{equation}   
\theta\ : \rho\ \otimes_{_{\sigma}}\ \tilde F\xrightarrow{\ \ \simeq\ \ }\ F  
\label{eq:LHS}  
\end{equation}   
of absolute $N$-dimensional theories. The LHS of \eqref{eq:LHS} defines the dimensional reduction, i.e. the sandwich we were referring to earlier. $\sigma$ needs only be a \emph{once-categorified} $N$-dimensional theory, whereas $\rho$ and $\tilde F$ are relative theories.

\begin{figure}[ht!]   
\begin{center}
\includegraphics[scale=0.9]{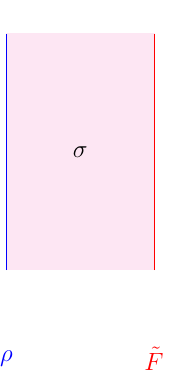}  
\label{fig:FrredMooreTeleman extended} 
\caption{\small Specifying the topological boundary conditions turns a relative QFT into an absolute one.}  
\end{center} 
\end{figure} 

Gauge theories are examples of relative theories, \cite{Freed:2012bs}.

More generally, \emph{relative theories}  are models whose fields form a fibration

\begin{equation}  
{\cal F}\ \longrightarrow\  {\cal M}\ \xrightarrow{\ \ \pi\ \ }\ {\cal B} 
\end{equation}   
where ${\cal B}$ and ${\cal F}$ denote the base and the fiber, respectively. Relative fields are the fibers of $\pi$. The path integral over the base reduces a finite sum. For the theory to be absolute, the same should hold for ${\cal F}$ as well. 

As originally outlined in \cite{Freed:2012bs}, \ QFT $F$ relative to $\alpha$ is a homomorphism 

\begin{equation}   
\tilde F:\ \tau_{_{\le n}}\alpha   \ \rightarrow\ \mathbf{1}\  
\end{equation}   
where $\alpha$ is the bulk theory w.r.t. $F$.

Let $G$ be a finite group. A classifying space $BG$ is derived   from a contractible topological space $EG$ equipped with a free $G$-action by taking the quotient. The homotopy type of $BG$ is independent of choices. If $X$ is a topological space equipped with a $G$-action, then the Borel construction is the total space of the fiber bundle.

\begin{equation}  
\begin{aligned}   
&X_{_{G}}\ \equiv\ EG\ \times_{_{G}}\ X  \\   
&\\    
&\downarrow\ \ \pi\\ 
&\\ 
&BG   
\end{aligned}   
\end{equation}  
with fiber $X$. If $*\in BG$ is a point, and we choose a basepoint in the $G$-orbit in $EG$ labelled by $*$, then the fiber $\pi^{^-1}(*)$ is canonically identified with $X$. The abstract group symmetry data is the pair $(BG,*)$, and a realisation of the symmetry $(BG,*)$ on $X$ is a fiber bundle over $BG$ together with an identification of the fiber over $*\in BG$ with $X$.

Following \cite{Freed:2012bs}, we outline two preliminary examples of relative field theories:

\begin{enumerate}  

\item Sigma models. 

\item Gauge theories.  

\end{enumerate}

\subsection*{1. Sigma models}

Data: $G$ a finite group and $M$ a smooth manifold with a free left-acting $G$ and quotient $\bar M$ (equivalently the gauged $\sigma$ model on $M$).

Denoting with $BG(X)$ the collection of principal $G$-bundles over $X$, 

A principal $G$-bundle is a covering space $P\rightarrow X$ and a free-$G$ action on $P$ such that $P\rightarrow X$ is a quotient map for the $G$-action. We denote the collection of principal $G$-bundles over $X$ with $BG(X)$, with the latter being a grupoid, and, as such, is a category in which every morphism is invertible. 

A \emph{symmetry}

\begin{equation}  
\varphi: (P\ \rightarrow\ X)\ \rightarrow   \ (P^{\prime}\ \rightarrow\ X)
\end{equation}  is a diffeomorphism 

\begin{equation}  
\varphi: P\ \rightarrow\ P^{\prime}  
\end{equation}  
commuting with $G$ and covering id$_{_{X}}$. The automorphism group of $P\ \rightarrow X$ is the group of gauge transformations. 

The path-integral over $BG(Y)$ is an integral over the equivalence classes of $G$-bundles. 

Fixing $(P\ \rightarrow\ X)\ \in\ BG(X)$, a relative field over $(P\ \rightarrow\ X)$ is a pair $(f,\theta)$ consisting of a smooth map $f: X\rightarrow \bar M$ and an isomorphism $\theta:(P\ \rightarrow\ X)\rightarrow f^*(M\ \rightarrow\ \bar M)$ of $G$-bundles over $X$. Given that relative fields feature no automorphisms, they are \emph{rigid}, and therefore constitute a space rather than a category. 

Given a pair of theories, $(\alpha, F)$, with $\alpha$ topological and defined on all manifolds, and $F$ a Riemannian theory, using the knowledge of the $\sigma$-model $F$ to predict the structure of $\alpha$. 

The relative path-integral $F(X)$ on a closed $D$-dimensional manifold is an integral over relative fields, hence a function   

\begin{equation}  
F(X): BG(X)\ \rightarrow \ \mathbb{C}  
\label{eq:neqFX}   
\end{equation}   

As such, $F(X)$ is invariant under symmetries of $BG(X)$, and is therefore expressed as a function on equivalence classes, $H^{^{1}}(X;G)$, with the latter denoting the isomorphic classes if principal $G$-bundles over $X$. From this, \eqref{eq:neqFX} can be re-expressed as 

\begin{equation}  
F(X): H^{^{1}}(X;G)\ \rightarrow \ \mathbb{C}  
\label{eq:neqFX1}   
\end{equation}   

If $F$ is to be a QFT relative to a $D+1$-dimensional bulk $\alpha$, in the sense that 

\begin{equation}  
F(X):\ \mathbb{C}\ \rightarrow\ \alpha(X)   
\end{equation}  
then, $F(X)$ can be identified as an element of the vector space $\alpha(X)$, leading to the conclusion that $\alpha(X)$ id the free vector space on the finite set $H^{^{1}}(X;G)$.  

Next, considering a closed  $D-1$-dimensional manifold, $Y$, the relative canonical quantisation of $F(Y)$ is obtained by performing the quantisation on the fibers of the map 

\begin{equation}  
\text{Map}(Y,\bar M)\ \rightarrow\ BG(Y)  
\end{equation}    
such that the vector bundle is 

\begin{equation}  
F(Y)\ \rightarrow\ BG(Y)  
\label{eq:equivbund}
\end{equation}

\begin{figure}[ht!]   
\begin{center}
\includegraphics[scale=0.5]{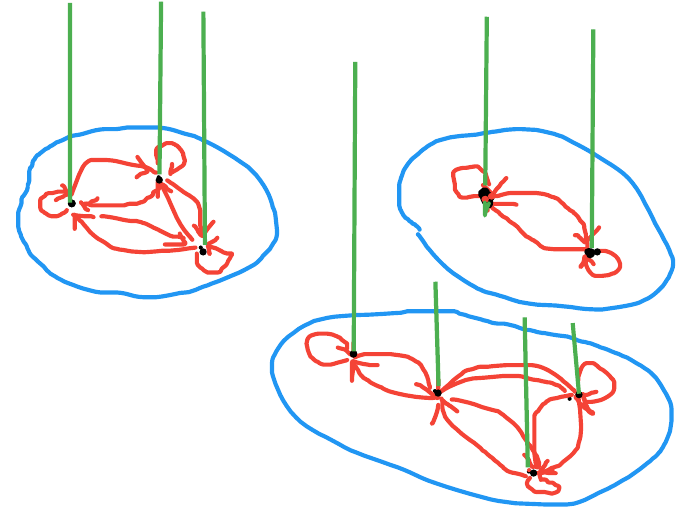} 
\label{fig:fiber}   
\caption{\small A depiction of the vector bundle.}
\end{center} 
\end{figure} 

The bundles are grouped into isomorphism classes. Points featuring in each isomorphism class denote the $G$-bundles $Q\rightarrow Y$, and the arrows the isomorphisms between the $G-bundles$. Selecting a representative bundle, $Q\rightarrow Y$, in each isomorphism class, the grupoid $BG(Y)$ corresponds to a simpler grupoid with a finite set of $H^{^{1}}(Y;G)$ and no morphisms between them.

If $m\ \in H^{^{1}}(Y;G)$ is the class of a $G$-bundle $Q\rightarrow Y$, then the automorphism group of $m$ is the group of gauge transformations of $Q\rightarrow Y$, and the equivariant bundle, \eqref{eq:equivbund}. decomposes into topological vector spaces indexed by pairs $(m,e)$ in which $m\ \in\ H^{^{1}}(Y;G)$ and $e$ is a complex irreducible representation of the automorphism group of $m$. The partition function on $Y$   

\begin{equation} 
F(Y):\ \mathbf{1}(Y)\ \rightarrow \ \alpha(Y)  
\end{equation}  
may be identified with an object belonging to the linear category $\alpha(Y)$, leading to the prediction that $\alpha(Y)$ is the free Vect$_{_{top}}$-module with basis pairs $(m,e)$.  

For the case in which $X$ is a 1D theory, namely $X\equiv\ S^{^{1}}$, $H^{^{1}}(S^{^{1}};G)$ is the set of \emph{conjugacy classes} in $G$. If $Y\equiv$pt, then $H^{^{1}}(\text{pt};G)$ has a single element, representing the trivial bundle with automorphism group $G$. Consequently, $\alpha(\text{pt})$ defines the category of representations of $G$. 

In this case, $\alpha$ is the finite 2D gauge theory with gauge group $G$, and can be defined on any manifold of $D\le 2$ by a finite path-integral. Given that the the theory has a vanishing Lagrangian, the partition function on a closed 2D manifold, $M_{_{2}}$, reduces to the weighted counting of $G$-bundles    

\begin{equation}  
\alpha(M_{_{2}})  
\ 
=   
\   
\sum_{[R\rightarrow\ M_{_{2}}]}\ \frac{1}{\ |\text{Aut}(R\rightarrow M_{_{2}})|\ }  
\end{equation}   

The finite path integral over $X\equiv S^{^{1}}$ is the space of invariant sections of the trivial equivariant line bundle $BG(S^{^{1}})$. On $Y\equiv$ pt, instead, the exponential of the action has constant value in the linear category Vect$_{_{top}}$, hence the path integral defines the subcategory of invariants under $G$, namely the category of representations of $G$.   

\subsection*{From relative to absolute}

An \emph{absolute} theory can be recovered from a relative one under gauging by a finite subgroup $G^{^{\prime}}\ \subset\ G$. In such case, the $\sigma$-model on a closed $D$-dimensional manifold can be expressed in terms of the relative theory as follows

\begin{equation}  
f_{_{G^{\prime}}}(X)   
\ 
=   
\   
\sum_{m^{\prime}\ \in\  H^{^{1}}(X;G^{\prime})}\ \frac{1}{\ |Z_{_{G^{\prime}}}(m^{\prime})|\ }\ F(X;\ m^{\prime})  
\end{equation} 
where the sum runs over the equivalence classes of $G^{\prime}$-bundles $P\rightarrow X$, and $Z_{_{G^{\prime}}}(m^{\prime})$ denotes the automorphism group of a representative of the equivalence class $m^{\prime}$. 

Following similar arguments, the quantum topological vector space on a closed $D-1$-dimensional manifold $Y$ can be expressed as

\begin{equation}  
f_{_{G^{\prime}}}(Y)   
\ 
=   
\   
\bigoplus_{m^{\prime}\ \equiv\ [Q^{\prime}\rightarrow Y]\in\  H^{^{1}}(X;G^{\prime})}\  F(Y;\ Q\rightarrow Y)^{^{\text{Aut}(Q^{\prime}\rightarrow Y)}}  
\end{equation}

\subsection*{2. Gauge theories}

A very similar analysis holds for the case of gauge theories, with all the fields' categorical levels enhanced by 1 unit, due to the presence of an additional symmetry. 

In this case, the theory is specified by the following data: a Lie algebra, $\mathfrak{g}$, a covering homomorphism $G\rightarrow\ \tilde G$ of compact connected Lie groups with kernel $G_{_{o}}$ a finite central abelian subgroup of $G$. For example, if $G_{_{o}}\ \subset\ G$ and $\tilde G\equiv G\diagdown G_{_{o}}$ is the adjoint group, this gives a canonical choice of data associated to a compact semisimple Lie algebra. 

Assuming $\tilde P\rightarrow X$ denotes a principal $\tilde G$ bundle, the obstruction to lifting to a principal $G$-bundle is measured by a $G_{_{o}}-gerbe$,  ${\cal G}(\tilde P)\rightarrow X$, and with $B^{^{2}}G_{_{o}}(X)$ the collection of $G_{_{o}}$-gerbes over $X$. $B^{^{2}}G_{_{o}}(X)$ is a 1-category. Its equivalence classes constitute the cohomology group $H^{^{2}}(X;G_{_{o}})$. The group of equivalence classes of automorphisms between objects is $H^{^{1}}(X;G_{_{o}})$, and the group of automorphisms of automorphisms is $H^{^{0}}(X;G_{_{o}})$. 

For gauge theories, a relative field can be defined as $({\cal G}\rightarrow X)\ \in\ B^{^{2}}G_{_{o}}(X)$. A relative field over $({\cal G}\rightarrow X)$ is a pair $(\tilde\Theta, \theta)$ consisting of a $\tilde G$-connection $\tilde \Theta$ and an isomorphism $\theta: {\cal G}\rightarrow{\cal G}(\tilde\Theta)$ of $G_{_{o}}$-gerbes. $(\tilde\Theta, \theta)$ form an ordinary grupoid, with no automorphisms between automorphisms. The fields are $\tilde G$-connections with $G$-gauge transformations.

The quantum theories $(\alpha, F)$ that can be built from these fields are a $D+1$ dimensional topological theory and a $D$-dimensional relative theory $F$. $\alpha$ is now defined as a path-integral over the $G_{_{o}}$-gerbes, specifically  

\begin{equation}  
\alpha(M_{_{2}})  
\ 
\equiv 
\    
\frac{\ |H^{^{2}}(X;G_{_{o}})| | H^{^{o}}(X;G_{_{o}})|\ }{|H^{^{1}}(X;G_{_{o}})| }     
\end{equation}  

Consequently, $\alpha(X)$ is the vector space of complex-valued functions on $B^{^{2}}(X)$. The relative theory $F$ is an element of $\alpha(X)$ since  

\begin{equation}
F(X)\equiv H^{^{2}}(X;G_{_{o}})\ \rightarrow \ \mathbb{C}  
\label{eq:neqFX1}   
\end{equation} 

 $\alpha(Y)$, instead, defines the linear category of vector bundles over $B^{^{2}}G_{_{o}}(Y)$. The relative theory $F$ determines a particular vector bundle $F(Y)\rightarrow B^{^{2}}G_{_{o}}(Y)$. 

 For any finite abelian group $A$, its Pontryagin dual is defined as follows  

 \begin{equation}  
 A^{^{\text{V}}}\ \overset{def.}{=}\ \text{Hom}(A,G_{_{o}})   
 \end{equation}  

 Once having chosen the basepoints, $F(Y)\rightarrow B^{^{2}}G_{_{o}}(Y)$ determines the topological vector spaces $F(Y;m,e)$ for $m\in H^{^{^2}}(Y;G_{_{o}}), e\in H^{^{1}}(Y;G_{_{o}})^{^{\text{V}}}$, where $(m,e)$ denote the classes of discrete magnetic and electric fluxes. 

\subsection*{From relative to absolute}

 Once more, an absolute gauge theory can be obtained by gauging by a subgroup of $G$, leading to a relic gauge group $G\diagdown G_{_{o}}^{\prime}$, $G_{_{o}}^{\prime}\subset G_{_{o}}$, and its expression reads

 \begin{equation}  
f_{_{G_{_{o}}^{\prime}}}(X)   
\ 
=   
\   
\sum_{m^{\prime}\ \in\  H^{^{2}}(X;G_{_{o}})}\ \frac{\ |H^{^{0}}(X;G_{_{o}}^{\prime})|\ }{\ |H^{^{1}}(X;G_{_{o}}^{\prime})|\ }\ F(X;\ m^{\prime}).  
\end{equation}

\begin{figure}[ht!]  
\begin{center}
\includegraphics[scale=0.9]{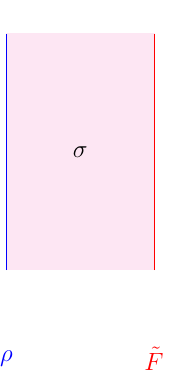} 
\ \ \ \ \ \ \ \ 
\includegraphics[scale=0.9]{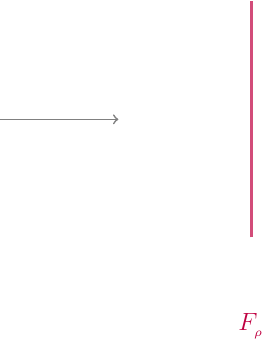}  
\caption{\small The Freed-Moore-Teleman setup, with $\tilde F$ denoting a relative QFT. Specifying the topological data $(\sigma,\rho)$, the resulting theory, $\tilde F_{_{\rho}}$ is absolute.}
\label{fig:FMT}    
\end{center}  
\end{figure}

In the formulation of \cite{Freed:2012bs}, a \emph{relative} field theory, $\tilde F$, requires additional topological data in order to be fully specified. Such data is encoded in a pair $(\sigma,\rho)$, referred to as \emph{quiche}. $\sigma$ is the \emph{symmetry topological field theory} (SymTFT), whereas $\rho$ geometrises the choice of boundary conditions for the fields defining the relative theory $\tilde F$. The overall system, depicted in figure \ref{fig:FMT}, gives rise to an \emph{absolute} QFT, $F_{_{\rho}}$.

In mathematical terms, the description outlined above can be formulated in terms of bordism in the following way. Fixing $N\in\mathbb{Z}^{^{\ge 0}}$, then a quiche is a pair $(\sigma, \rho)$ in which $\sigma:\text{Bord}_{_{N+1}}(F)\rightarrow{\cal C}$ is an $N+1$-dimensional TFT and $\rho$ is a right topological $\sigma$-module. The quiche is $N$-dimensional, hence it shares the same dimensionality as the theory on which it acts. Let $F$ be an $N$-dimensional field theory. A $(\sigma,\rho)$-\emph{module structure} on $F$ is a pair $(\tilde F, \theta)$, in which $\tilde F$ is a left $\sigma$-module and $\theta$ is an isomorphism 

\begin{equation}   
\theta\ : \rho\ \otimes_{_{\sigma}}\ \tilde F\xrightarrow{\ \ \simeq\ \ }\ F_{_{\rho}},  
\label{eq:LHS}  
\end{equation}   
of absolute $N$-dimensional theories, with \eqref{eq:LHS} defining the dimensional reduction leading to the absolute theory. $\sigma$ needs only be a \emph{once-categorified} $N$-dimensional theory, whereas $\rho$ and $\tilde F$ are relative theories.

\begin{figure}[ht!]   
\begin{center}
\includegraphics[scale=0.7]{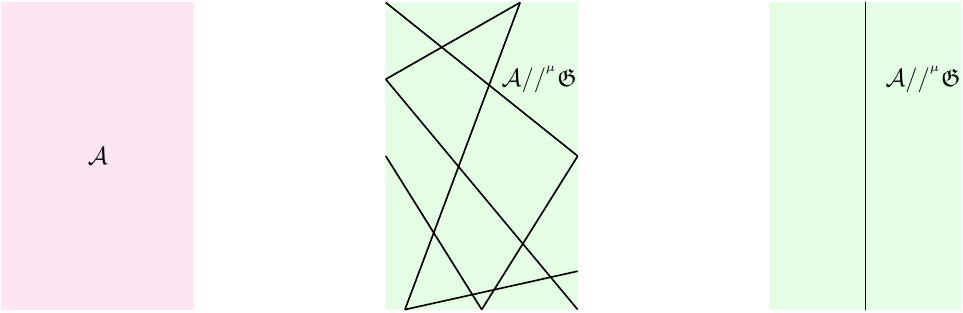}     
\caption{\small Gauging corresponds to condensing an algebra in a TFT. Idempotency ensures the resulting theory can be effectively thought of as featuring a unique defect, as shown on the RHS.}
\label{fig:codensationTJF}  
\end{center} 
\end{figure}

 Having explained the notion of relative QFTs, we now provide some further explanation for what gauging a categorical structure actually means. In doing so, we refer to the work of many experts in the field, and, in particular \cite{TJF}. As explained in such reference, for any fusion n-category $\mathfrak{G}$, any fiber functor

\begin{equation} 
{\cal F}:\ \mathfrak{G}\ \rightarrow\ \text{nVec},  
\label{eq:functor}   
\end{equation}   
selects nVec as the image of a condensation algebra living in $\mathfrak{G}$, corresponding to a projection on the identity. The gauging process, can therefore be defined as a map

\begin{equation} 
\mu:\ \mathfrak{G}\ \rightarrow \ {\cal A}_{_C}, 
\label{eq:mu}   
\end{equation}   
with ${\cal A}_{_C}$ the algebra of invertible topological operators in $\mathfrak{G}$. Given \eqref{eq:mu}, the norm element

\begin{equation} 
N\overset{def}{=}\bigoplus_{g\in\mathfrak{G}}\ \mu(g).  
\label{eq:idemp}   
\end{equation}    
carries the structure of an n-categorical idempotent, also known as \emph{condensation algebra}, depicted in black in figure \ref{fig:codensationTJF}. The requirement for \eqref{eq:idemp} to be a higher-idempotent is needed to ensure the flooding doesn't depend on the specific features of the network being adopted to perform the gauging. The algebra of topological operators that are left are denoted by ${\cal A}//^{^{\mu}}\mathfrak{G}$. The equivalence of the second and third picture from the left in figure \ref{fig:codensationTJF} follows from $N$ being idempotent. This pattern emerges, for example, when gauging the 5D SymTFT of class ${\cal S}$ theories, with the objects of the 2-category in question being Wilson surfaces charged under the 1-form symmetry being gauged.

\section*{Key point} 

The aim of the analysis carried out in the present and following section is that of explaining how ring homologies are actually related to higher-categorical structures in the SymTFT construction of \cite{Freed:2022qnc}.

The key points to bare in mind for the following section can be summarised in terms of two mutually related points:  

\begin{enumerate}  

\item The need to keep track of the identity of the gaugeable algebra. Indeed, the moment map \eqref{eq:mu} is essentially providing a redefinition of the identity once the gauging has been performed\footnote{Specifically, the identity element in the pre-gauged algebra, $\mathbf{1}$, is the composite algebra ${\cal A}_{_C}$ with respect to which the gauging has been performed.}. 

\item The embedding of the gauged algebra. 

\end{enumerate} 

By the end of section \ref{sec:4} it will become clear how both points are related to abelianisation, \cite{Dimofte:2018abu,Bullimore:2015lsa}, and how this is crucial for accessing the mathematical structure of a given quiven QFT.

\section{Quivers}    \label{sec:5}    

Equipped with the tools revised in the previous sections, we are now ready to begin the analysis outlined in the roadmap illustrated in figure \ref{fig:roadmap}. The top entry of our diagram reads an example of a 2-category, namely that of Moore-Tachikawa varieties\footnote{To facilitate the reader, we have added an appendix with further details to be read, albeit we highly recommend to go back to the original paper \cite{Moore:2011ee} for a more thorough overview.}. The reason why we are referring to them as our main example is their relation to Coulomb branches of star-shaped quivers, \cite{Dimofte:2018abu}, to which we can apply the analysis of section \ref{sec:4}. The setups we are dealing with are highly supersymmetric, thereby describing theories that are very far from the SM of particle physics, even in its minimally supersymmetric realisation (MSSM). However, the purpose of this article is that of highlighting features in categorical structures of supersymmetric gauge theories that could come to aid in furthering the understanding of extensions of the SM to possible completions in unifying theories, in turn descending from a superymmetric parent setting, for the reasons outlined in section \ref{sec:2}. Furthermore, it is important to remark that, the quiver description of the SM and its extensions is an essential step towards embedding them in any string theory setup, and we refer the reader to some key works on this regard, such as \cite{Cicoli:2021dhg, Uranga}.

In section \ref{sec:4}, we have gathered sufficient evidence supporting the utility of applying category theory for describing simple examples of QFT models. 
We now switch gears, and turn to a more algebro-geometric analysis of the setup outlined so far. In doing so, we will mostly be relying on \cite{Braverman:2017ofm,Braverman:2016wma,Dimofte:2018abu,Bullimore:2015lsa}, shedding new light and emphasising techniques that should be further developed and applied to Beyond the Standard Model setups\footnote{More details in this regard will appear in an upcoming work by the same author, \cite{VP}.}. 

The present section is structured as follows:  

\begin{enumerate}  

\item  At first, we overview some key features of equivariant cohomology and the convolution product, both within the context of quiver gauge theories, \cite{Braverman:2017ofm,Braverman:2016wma}

\item Then, we introduce Moore-Tachikawa varieties and their 2D TFT counterpart\footnote{For completeness, we provide some further details in appendix \ref{sec:mt}.}, connecting categorical structures and quiver varieties, \cite{Moore:2011ee}. Making use of \cite{Dimofte:2018abu}, we emphasise how and when 3D mirror symmetry manifests from a categorical perspective, specifically in terms of bordisms, making use of the Kostant-Whittaker symplectic reduction, \cite{Braverman:2017ofm}.

\item  Last but not least, we close the section highlighting the importance of abelianisation for embedding the ring homology of the theory in question, explaining how this is in turn related to (un)gauging the categorical SymTFT introduced in section \ref{sec:4}.

\end{enumerate}  

\subsection{Equivariant cohomology, fixed point localisation and convolution product}

Equivariant cohomology, \cite{BFM}, also known as Borel cohomology, is a cohomology theory from algebraic topology which applies to topological spaces with a group action. It can be viewed as a generalisation of group cohomology. More explicitly, the equivariant cohomology ring of a space $X$ with action of a topological group $G$ is defined as the ordinary cohomology ring with coefficient ring $\Lambda$ of the homotopy quotient $EG\times_{_G}X$  

\begin{equation}  
H^{^{\bullet}}_{_G}(X;\Lambda)\ =\ H^{^{\bullet}}\left(EG\times_{_G} X;\Lambda\right).    
\end{equation}  

\subsection*{The Kirwan map}  

The Kirwan map, first introduced in \cite{FK}, states 

\begin{equation}  
H^{^{\bullet}}_{_G}(M)\ \rightarrow\ H^{^{\bullet}}\left(M//_{_p} G\right), 
\end{equation}  
where $M$ is a Hamiltonian $G$-space, i.e. a symplectic manifold acted on by a Lie group $G$ with a moment map 

\begin{equation}     
\mu:\ M\ \rightarrow\ \mathfrak{g}^{*}.  
\end{equation}   

$H^{^{\bullet}}(M)$ is an equivariant cohomology ring of $M$, i.e. the cohomology ring of the homotopy quotient $EG\times_{_G}M$ of $M$ by $G$.

\begin{equation}  
M//_{_p}G\ =\ \mu^{^{-1}}(p)/G  
\end{equation}   
is the symplectic quotient of $M$ by $G$ at a regular central value $p\in Z(\mathfrak{g}^{*})$ of $\mu$. It is defined as the map of equivariant cohomology induced by the inclusion 

\begin{equation}    
\mu^{^{-1}}(p)\ \hookrightarrow\ M   
\end{equation}
followed by the canonical isomorphism  

\begin{equation}  
H^{^{\bullet}}_{_G}\left(\mu^{^{-1}}(p)\right)\ =\ H^{^{\bullet}}\left(M//_{_p} G\right).
\end{equation} 

\subsection*{Fixed point localisation and convolution product}   

In BFN, the embedding of algebras is performed under the so called \emph{convolution product}. To see how this is defined, we start off with the convolution diagram for the affine Grassmannian, \cite{Braverman:2017ofm, Braverman:2016wma} 

\begin{equation}  
\text{Gr}_{_G}\times\text{Gr}_{_G}\ \overset{p}{\leftarrow}\ G_{_{\cal K}}\ \times\ \text{Gr}_{_G}\ \overset{q}{\rightarrow}\ \text{Gr}_{_G}\tilde\times\ \text{Gr}_{_G}\ \overset{m}{\rightarrow}\ \text{Gr}_{_G},   
\end{equation}   
where $p,q$ are projections, $m$ is a multiplication, and

\begin{equation}  
\text{Gr}_{_G}\tilde\times\ \text{Gr}_{_G}\ =\ G_{_{\cal K}}\ \times_{_{G_{_{\cal O}}}}\ \text{Gr}_{_G}. 
\label{eq:convoldiagr}     
\end{equation}

Hence, given any two $G_{_{\cal O}}$-equivariant perverse sheaves ${\cal A}_{_1}, {\cal A}_{_2}$, the pullback  

\begin{equation}  
p^{^*}\left({\cal A}_{_1}\boxtimes {\cal A}_{_2}\right) \ \equiv\ {\cal A}_{_1}\ \tilde\boxtimes\ {\cal A}_{_2}   
\end{equation}  
descends to $\text{Gr}_{_G}\ \tilde\times\ \text{Gr}_{_G}$ by equivariance, and we can define the convolution product as follows  

\begin{equation}  
{\cal A}_{_1}\star {\cal A}_{_2}  \ =\ m_{_*}\left({\cal A}_{_1}\ \tilde\boxtimes\ {\cal A}_{_2}\right).  
\end{equation} 

The latter defines a symmetric monoidal structure on the category of $G_{_{\cal O}}$-equivariant perverse sheaves on $\text{Gr}_{_G}$, and is equivariant to the monoidal category of finite-dimensional representations of the Langlands dual of $G$.   

\eqref{eq:convoldiagr} is the convolution diagram for the case in which $G$ is a complex reductive group and representation $\mathbf{N}=0$. For arbitrary $\mathbf{N}$, we can generalise the convolution diagram, achieving a noncommutative algebraic structure. In order to do so, we define a triple ${\cal R}=({\cal P},\varphi, s)$, where ${\cal P}$ is a $G$-bundle on a formal disk $D=\text{Spec}\ \mathbf{C}[[z]]$, $\varphi$ is its trivialisation over the punctured disk $D^{^*}=\text{Spec}\ \mathbf{C}((z))$, and $s$ is a section of the associated vector bundle ${\cal P}\times_{_G}\mathbf{N}$ such that it is sent to a regular section of the trivial bundle under $\varphi$.

\subsection*{Coulomb branches} 

Given a complex vector space, $V$, on a connected reductive algebraic group, $G$, with a fixed faithful linear action on $V$, the \emph{Higgs branch} is defined as follows  

\begin{equation}  
{\cal M}_{_H}\ =\ \mu^{^{-1}}(0)\ //\ G\ =\ \text{Spec}\left(\mathbf{C}\left[\mu^{^{-1}}(0)\right]^{^G}\right)  
\end{equation}   
with 

\begin{equation}   
\mu:\ T^{^*}V\ \rightarrow \ \mathfrak{g}  
\end{equation}  
denoting the moment map, whereas the \emph{Coulomb branch}, ${\cal M}_{_C}$, is the spectrum of a ring constructed as a convolution algebra in the homology of the affine Grassmannian, where the choice of the representation $V$ is incorporated as certain quantum corrections to convolution in homology, which are kept track of by an auxiliary vector bundle, \cite{Braverman:2017ofm, Braverman:2016wma}. ${\cal M}_{_H}$ and ${\cal M}_{_C}$ are conjectured to be symplectic duals to each other, with the most important relation in between them being Koszul duality between the categories of the quantised varieties, \cite{Webster:2016rhh}.

Let $G$ be a simply-connected complex semisimple group, and let $G_{_{ad}}$ be its adjoint form. The group $G_{_{ad}}$ acts on $G$ by conjugation, and $G$ contains a transverse slice $\Sigma$ for this action, which was first introduced by Steinberg, \cite{stein}. The resulting multiplicative universal centraliser is the smooth affine variety   

\begin{equation}    
Z_{_{G}}\ =\ \left\{(a,h)\ \in\ G_{_{ad}}\times\Sigma\ \bigg|\ a\in G_{_{ad}}^{^h}\right\}.      
\end{equation}   

When $G$ is simply-laced, Bezrukavnikov, Finkelberg, and Mirkovíc, \cite{BFM}, showed that its coordinate ring is isomorphic to the equivariant K-theory of the affine Grassmannian of the Langlands dual group $G^{^{\text{V}}}$, therefore, in this case, $Z_{_{G}}$ is an example of a Coulomb branch as defined by BFN \cite{Braverman:2017ofm, Braverman:2016wma}.

\subsection{Moore-Tachikawa varieties}  

We now turn to the specific case of interest to us, namely Moore-Tachikawa varieties, and their relation with the work of \cite{Freed:2022qnc,Dimofte:2018abu,Teleman:2014jaa,Bullimore:2015lsa}. First and foremost, we will briefly overview how the Higgs and Coulomb branches arise in this setup.

Given a reductive connected group, $G$, there is a functor defined as follows   

\begin{equation}  
\eta_{_{G_{_{\mathbf{C}}}}}:\ \ \text{Bo}_{_2}\ \rightarrow\ {\cal C} 
\end{equation} 
where ${\cal C}$ is a hyperbolic symplectic manifold, whose objects are affine algebraic groups, and whose morphisms 

\begin{equation}   
\text{Hom}_{_{\cal C}}(G_{_1}, G_{_2})  
\end{equation}    
are isomorphism classes of symplectic\footnote{By this we mean affine, Poisson, operationally symplectic, with a Hamiltonian chosen moment map.} varieties with Hamiltonian $G_{_1}\times G_{_2}$-action. $\eta_{_{G_{_{\mathbf{C}}}}}$ is fully specified by its action on objects and morphisms, namely

\begin{equation}   
\begin{aligned}
&\eta_{_{G_{_{\mathbf{C}}}}}\left(S^{^1}\right)\ \mapsto\ G\ \ \ \ \ \ \ \ ,  \ \ \ \ \ \ \ \ \eta_{_{G_{_{\mathbf{C}}}}}\left(S^{^2}\right)\ \mapsto\ Z_{_G}\\  
&\\
&\ \ \ \ \ \ \ \ \ \ \ \ \ \ \ \ \eta_{_{G_{_{\mathbf{C}}}}}\left(S^{^1}\times S^{^1}\right)\ \mapsto\ T^{^*}G\\   
\end{aligned}   
\label{eq:maps}   
\end{equation}  
where $Z_{_G}$ is the \emph{universal centraliser}, consisting of all the pairs $(x,g)$ with  $x\in\mathfrak{g}$, $\mathfrak{g}$ Lie, regular, $g\in G$ and $ad(x)_{_{\mathfrak{g}}}=x$. Given $K\subset\mathfrak{g}$ a Kostant slice, and $(e, h,f)$ a principal $\mathfrak{sl}_{_2}$-triple in $\mathfrak{g}$, the Kostant slice can be rewritten as follows  

\begin{equation}  
K=e+Z(f)   
\end{equation}  
with $Z(f)$ the centraliser of $f$ in $\mathfrak{g}$, $\mathfrak{g}\simeq \mathfrak{g}^{^*}$. For a given Kostant slice, the universal centraliser can therefore be redefined as follows  

\begin{equation}  
Z_{_G}=\left\{(g, x)|g\in G, x\in K, \text{ad}_{_{\mathfrak{g}}}(x)=x\right\}.  
\end{equation}  

What we have outlined so far can be suitably adapted to the case of interest to us, namely when $\Sigma$ is a Riemann surface whose boundary has $n$ components, from which one gets a symplectic variety $X$ with a $G^{^n}$-action. This symplectic variety, $X$, is often times referred to as the \emph{Higgs branch}.

\subsubsection{The Kostant-Whittaker symplectic reduction}

The action of the functor explicited in \eqref{eq:maps}, however, is not exhaustive. Indeed, there is one more piece of information that is needed in order for the 2D TFT to be fully determined, and this is its action on the bordism corresponding to the three-punctured sphere.  

To see how this relates to quivers, let us consider the case in which $X$ is a symplectic variety with $G$-action. Let $N\subset G$ be a maximal map, and 

\begin{equation}         
\chi:\ \ N\ \longrightarrow\ \ \mathbf{C}  
\end{equation}    
a non-degenerate homomorphism such that 

\begin{equation}  
N\ \bigg/\ [N,N]\ \simeq\ \bigoplus_{_{{\alpha}_{_i}}}\ \mathbf{C}_{_{\alpha_{_i}}},  
\end{equation}   
where $\alpha_{_i}$ denote the simple roots. Then, the Kostant-Whittaker reduction on the symplectic variety corresponds to the quotient   

\begin{equation}  
KW(X)\ =\ \mu^{^{-1}}_{_N}(\chi)\ \bigg/\ N,  
\end{equation}
and the universal centraliser becomes   

\begin{equation}  
Z_{_G}\ \simeq\ KW_{_{G\times G}}\left(T^{^*}G\right).
\end{equation}    

Furthermore, we can take 

\begin{equation}  
G\times K = KW_{_{G^{^\text{V}}}}\left(T^{^*}G\right)  \ \ \ \ \ , \ \ \ \ \ KW_{_G}(G\times K)\ =\ Z_{_G},  
\end{equation}  
meaning that filling in a hole on the sphere corresponds to performing a $KW_{_G}$ reduction. Note that this is a statement of 3D mirror symmetry, \cite{Intriligator:1996ex}.

    According to the Moore-Tachikawa conjecture, overviewed in slightly more detail in appendix \ref{sec:mt}, a 2D TFT is an example of a 2-categorical structure, that is the Alday-Gaiotto-Tachikawa dual of a supersymmetric 4D theory arising from dimensional reduction of 6D ${\cal N}=(2,0)$ SCFTs of type $A_{_{N-1}}$ on a sphere with $N$ maximal punctures. These theories, oftentimes denoted ${\cal T}_{_{N}}\left[\Sigma_{_{0,k}}\right]$, upon being further dimensionally reduced on a circle, ${\cal T}_{_{N}}\left[\Sigma_{_{0,k}}\times S^{^1}\right]$, are known to be related to Sicilian 3D ${\cal N}=4$ theories, the so-called star-shaped quivers, upon taking the $\underset{S^{^1}\rightarrow0}{\lim}$, \cite{Dimofte:2018abu}.  

In \cite{Cremonesi:2014vla}, the authors calculated the Hilbert series of the chiral ring 

\begin{equation}         
\mathbf{C}\left[{\cal M}_{_{C}}\right]\ \simeq\ \mathbf{C}\left[{\cal M}_{_{H}}^{^{4D}}\right]  
\end{equation}  
exaclty with the intention of obtaining new information about the structure of the Higgs branches of the Higgs branch of the 4D theory, ${\cal M}_{_{H}}^{^{4D}}$, via a direct analysis of the corresponding Coulomb branches ${\cal M}_{_{C}}$ of ${\cal T}_{_{N,k}}$, for general $N$ and $k$. In addition to this, our work wishes to emphasise the importance of the calculation of the Hilbert series together with the correct identification of the ring elements and relations in between them, which plays a crucial role in developing connection with Particle Physics and calculations therein. In achieving the second objective, a crucial role is played by the so-called \emph{abelianisation} procedure, first introduced by \cite{Bullimore:2015lsa}, and subsequently emphasised also in \cite{Dimofte:2018abu}.

\subsection{Abelianisation}  

The basic idea of abelianisation is to embed the Coulomb branch chiral ring $\mathbf{C}[{\cal M}_{_C}]$ of a non-abelian theory into a larger, semisimple abelian algebra, ${\cal A}$,   

\begin{equation}  
\boxed{\ \ \ \mathbf{C}[{\cal M}_C]\ \hookrightarrow\ {\cal A}\ \ \color{white}\bigg]}.    
\label{eq:emb}
\end{equation}

Essentially, abelianisation corresponds to fixed-point localisation in the equivariant homology, which we will briefly explain momentarily. The embedding \eqref{eq:emb} allows to:    

\begin{enumerate}    

\item  Verify relations among elements of $\mathbf{C}[{\cal M}_{_C}]$.

\item Identify the Poisson structure on $\mathbf{C}[{\cal M}_{_C}]$ and its deformation quantisation.  

\item Extend the algebra $\mathbf{C}[{\cal M}_{_C}]$ over twistor space, ultimately enabling to access the hyperk$\ddot{\text{a}}$hler structure on ${\cal M}_{_C}$.   

\end{enumerate}

In the context of BFN, this arises when considering a 3D ${\cal N}=4$ gauge theory on $\mathbb{C}\times\mathbb{R}$ spacetime with $\frac{1}{2}$-BPS boundary conditions ${\cal B}$ near spatial infinity on $\mathbb{C}$.

\begin{figure}[ht!]  
\begin{center}  
\includegraphics[scale=0.8]{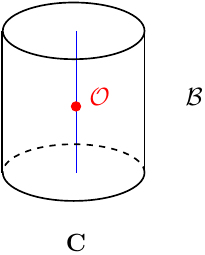}  \ \ \ \ \ \ \ \ \ \  \ \ \ \ \ \ \ \ \ \
\includegraphics[scale=0.8]{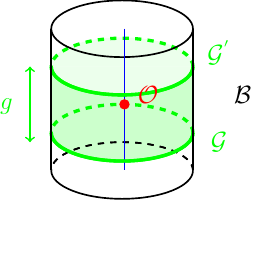} 
\caption{\small Picture reproduced from the work of \cite{Dimofte:2018abu}, encoding the key idea outlined in their treatment, and from which our work was in turn inspired.}  
\label{fig:cyl}   
\end{center} 
\end{figure}

Working in the cohomology of the Rozansky-Witten theory, ${\cal Q}_{_{RW}}$, the fields may be localised in the QM to sections of bundles that solve the ${\cal Q}_{_{RW}}$ BPS equations. This yields the gauged ${\cal N}=4$ QM on   

\begin{equation}  
{\cal M}_{_{[\mathbf{C}]}}=\left\{\ (E,X)\ \right\},  
\end{equation}  
where $E$ is a holomorphic $G_{_{\mathbf{C}}}$-bundle on $\mathbf{C}$, and $X$ is a holomorphic section of an associated $R$-bundle with gauge group, ${cal G}$ is the holomorphic $G_{_{\mathbf{C}}}$ gauge transformation on $\mathbf{C}$.  

The Hilbert space of the space of the QMs in the ${\cal Q}_{_{RW}}$ twist should be the ${\cal G}$-equivariant de Rham cohomology of ${\cal M}_{_{[\mathbf{C}]}}$. Turning on the Omega background further, corresponds to working equivariantly w.r.t. the $U(1)_{_{\epsilon}}$ spatial rotation group of $\mathbf{C}$, which is an ordinary symmetry of the moduli space, the Hilbert space is   

\begin{equation}   
{\cal H}\ \overset{def.}{=}\ H^{^{\bullet}}_{_{{\cal G}\times U(1)_{_{\epsilon}}}}\left({\cal M}_{_{[\mathbf{C}]}}\right).  
\end{equation}

${\cal M}_{_{[\mathbf{C}]}}$ is contractible to a point where $E$ is the trivial bundle and $X$ is the zero-section. ${\cal G}$ is contractible to $G$, therefore the Hilbert space reduces to   

\begin{equation}  
{\cal H}\ \simeq\ H^{^{\bullet}}_{_{{\cal G}\times U(1)_{_{\epsilon}}}}\left(\text{pt.}\right) =\mathbf{C}[\varphi,\epsilon]^{^W},   
\end{equation}    
namely Weyl-invariant polynomials in the equivariant weights $\varphi\in\mathfrak{t}_{_{\mathbf{C}}}$ and $\epsilon$.   

We can therefore write the BFN construction, which essentially reads as follows  

\begin{equation}  
\mathbf{C}_{_{\epsilon}}\left[{\cal M}_{_C}\right]\ \equiv\ H^{^{\bullet}}_{_{{\cal G}\times {\cal G}^{^{\prime}}\times U(1)_{_{\epsilon}}}}\left({\cal M}_{_{[\mathbf{C}\cup\mathbf{C}]}}\right)   \simeq H^{^{\bullet}}_{_{G\times U(1)_{_{\epsilon}}}}\left({\cal M}_{_{[\mathbf{C}\cup\mathbf{C}]}}\ \bigg /\ {\cal G}^{^{\prime}}\right),   
\label{eq:space}   
\end{equation}   
where ${\cal G,G}^{^{\prime}}$ are the groups of regular holomorphic gauge transformations on the top and bottom copies of $\mathbf{C}$, and $U(1)_{_{\epsilon}}$ is the usual spatial rotation group. Importantly, this space, \eqref{eq:space}, is non-contractible, and has highly nontrivial topology since it contains monopole operators. OPEs of local operators in the algebra $\mathbf{C}_{_{\epsilon}}[{\cal M}_{_{\epsilon}}]$ naturally correspond to the convolution product rather than the cup product.     

We are now ready to explain a particularly useful tool in the context of equivariant cohomology, namely fixed-point localisation. Letting $T\subset G$ denote the maximal torus, one finds that $T\times U(1)_{_{\epsilon}}$ fixed points of ${\cal M}_{_{[\mathbf{C}\cup\mathbf{C}]}}\bigg / {\cal G}^{^{\prime}}$ are isolated points described as pairs $(E, X), (E^{^{\prime}}, X^{^{\prime}})$ where $X=X^{^{\prime}}=0$ are zero sections, $E$ is trivial, and $E^{^{\prime}}$ is obtained from $E$ by a gauge transformation  

\begin{equation}  
g(z)=z^{^{A}}\ \ \ ,\ \ \ A\in\ \text{cochar}(G),  
\end{equation}  
meaning fixed points are labelled by cocharacters corresponding to ableian monopole operators. 

If ${\cal F}$ is the fixed point set of the $T\times U(1)_{_{\epsilon}}$ action on ${\cal M}_{_{[\mathbf{C}\cup\mathbf{C}]}}\bigg / {\cal G}^{^{\prime}}$, ${\cal F}\simeq\text{cochar}(G)$ is isomorphic to the cocharacter lattice. 

The equivariant cohomology of the fixed point set contains a copy of

\begin{equation}   
H^{^{\bullet}}_{_{T\times U(1)_{_{\epsilon}}}} (\text{pt})=\mathbf{C}[\varphi, \epsilon]  
\end{equation}  
for every point in ${\cal F}$, i.e.  

\begin{equation}   
H^{^{\bullet}}_{_{T\times U(1)_{_{\epsilon}}}} ({\cal F})\simeq\mathbf{C}\left[\varphi, \epsilon, \{v_{_{A}}\}_{_{A\ \in\ \text{cochar}(G)}}\right].  
\end{equation}  

In its localised version, one gets that 

\begin{equation}   
{\cal A}_{_{\epsilon}}\ \subset\ H^{^{\bullet}}_{_{T\times U(1)_{_{\epsilon}}}} ({\cal F})^{^{\text{loc}}}.   
\label{eq:aeht}   
\end{equation}  

The crucial difference in between the two sides in \eqref{eq:aeht} is that the localised cohomology indiscriminately inverts all weights, whereas in ${\cal A}_{_{\epsilon}}$ we only invert part of them. The localisation theorem provides the following algebraic embedding 

\begin{equation}  
\mathbf{C}_{_{\epsilon}}\left[{\cal M}_{_{[\mathbf{C}]}}\right]=H^{^{\bullet}}_{_{G\times U(1)_{_{\epsilon}}}}\left({\cal M}_{_{[\mathbf{C}\cup\mathbf{C}]}}\bigg /\ {\cal G}^{^{\prime}}\right)\ \hookrightarrow\ H^{^{\bullet}}_{_{T\times U(1)_{_{\epsilon}}}}\left({\cal F}\right)^{^{\text{loc}}},    
\label{eq:emb1}
\end{equation}  
which is the abelianisation we introduced in \eqref{eq:emb}. Importantly, the RHS of \eqref{eq:emb1} is the mother algebra we were looking for, the un-gauged algebra referred to in section \ref{sec:4}, and the key entity encoding essential information about the QFT in question. In conclusion to this section, we wish to highlight that, for any successful QFT to be fully specified, one should really be able to determine such an embedding. Any shortcomings in achieving it, should be taken as signalling a yet incomplete understanding of the underlying mathematical structure of the theory in question.

\section{Where to next?}   \label{sec:6}

With the discussion in section \ref{sec:4}, we have reached the bifurcation point in diagram \ref{fig:roadmap}, and showed the importance of abelianisation in geometric representation theory. Specifically, we highlighted the common ground between \cite{Freed:2022qnc,Dimofte:2018abu,Teleman:2014jaa}. 

As explained at the very beginning of our work, one of our key motivations is the need to deepen the mathematical toolbox adopted by Phenomenologists in answering open questions in Particle Physics. Recent developments in the Pheno community have started relying upon the calculation of the Hilbert series, most importantly \cite{Graf:2022rco,Bento:2023owf,Anisha:2019nzx}, successfully encoding key fundamental processes, such as the Higgs mechanism. The crucial point of our analysis, though, is that, albeit Hilbert series calculations are indeed very useful for keeping track of the degrees of freedom and symmetries of a given theory, in principle they are not enough to determine the exact operator content of the theory in question. 

On the other hand, ring homologies, on which Hilbert series are calculated, encode crucial properties of the QFT at hand, enabling a much deeper understanding of its underlying categorical structure. The key take-home message of this work is the crucial role played by abelianisation for identifying the underlying mathematical structure of a given QFT. This concept will be investigated further for the specific case of the SM and its extensions in future work, \cite{VP}. 

Additional complementary directions addressed by the same author will be developing ring homologies for Moore-Tachikawa varieties beyond categorical duality\footnote{Thereby following up from the analysis carried out in \cite{Pasquarella:2023ntw}.}, and understanding their implications for Koszul duality. We plan to report of any advancements in this regard in the near future.

\section{Conclusions and Outlook} 

This work is meant to be the first of a series of papers by the same author, continuing from past developments, bridging the gap in between the Pure Mathematics and the Theoretical Particle Physics. The purpose of this introductory article was mainly:  

\begin{itemize}  

\item   To illustrate (some) key fundamental problems in Theoretical Particle Physics necessitating further mathematical understanding.   

\item   Opening the scene to well-established techniques in representation theory and categorical algebraic geometry to help in this regard.    

\item   For those who are not experts in the field, serve as an oversimplified explanation of the powerful formalism put forward by BFN, and connecting it with Moore-Tachikawa varieties.   

\item Motivating Phenomenologists to engage more with Pure Mathematicians.

\end{itemize}

The main objective of this paper was that of explaining the higher categorical structures that are needed for describing the invariants associated to specific supersymmetric quiver gauge theories, with a particular focus on dualities and their mutual relations in terms of higher-categories.

The article was structured as follows: 

\begin{enumerate}

\item Section \ref{sec:2}  explained the motivations and outline of the present and forthcoming work by the same author, \cite{VP}. In doing so, we briefly introduced scattering amplitudes as crucial tools within the context of particle physics, explaining their connection with Hilbert series. We then briefly explained some of the key open questions in Theoretical Particle Physics necessitating further mathematical understanding, specifically the hierarchy problem, and the lightness of the Higgs mass (and its relation to Supersymmetry (SUSY)).

\item     Section \ref{sec:4}   is mostly a revision of \cite{Freed:2022qnc,Freed:2012bs}, and we used it at this stage for multiple reasons. Primarily because it allows to introduce gauge theories and Lagrangians in a mathematical language\footnote{Both of which are essential in the description of the SM and its extensions.}, which in turn is suitable to make connection with the higher-categorical Symmetry Topological Field Theory (SymTFT) prescription put forward by Freed, Moore, and Teleman, (FMT), \cite{Freed:2022qnc}, in the context of relative QFTs.  

\item     Section \ref{sec:5} was ultimately devoted to explaining in a simplified way the well-established BFN construction leading to ring homologies. The theories in question are 3D ${\cal N}=4$ supersymmetric quiver gauge theories. This section highlights the importance that categorical algebraic geometry plays in identifying ring homologies accounting for the complete spectrum of a given theory, therefore giving a complementary insight with respect to the Hilbert series alone. In so doing, we mostly highlighted the importance of \emph{abelianisation}.  In doing so we explicitly showed the relation in between the work of \cite{Bullimore:2015lsa}, that of Freed, Moore, and Teleman (FMT), \cite{Freed:2022qnc}, and Teleman's, \cite{Teleman:2014jaa}, which, to the best of our knowledge, has only previously been mentioned in \cite{Pasquarella:2023exd,Pasquarella:2023}. 

\item From the discussion of the previous section, the concluding section, \ref{sec:6}, paves the way to forthcoming work \cite{VP}, where a more detailed analysis with regard to the algebraic structure and Hilbert series of (Beyond) the Standard Model scenarios will be carried out. Complementarily to this, our objective is that of developing ring homologies for Moore-Tachikawa varieties beyond categorical duality, \cite{Pasquarella:2023ntw}, and understanding their implications for Koszul duality. We plan to report of any advancements in this regard in the near future.

\end{enumerate}

\section*{Acknowledgements}

I wish to thank the Cambridge Pheno Working Group for instructive weekly seminars, opening my perspectives towards new aspects of Quantum Field Theory.
This work is partially supported by the STFC Consolidated HEP theory grant ST/T000694/1 through DAMTP.

\appendix

\section{Algebraic varieties: some more background details}  

\subsection*{Riemann surfaces}

The geodesic structure of a Riemannian manifold is described by the Hamilton-Jacobi equations. As previously mentioned, the $M$ and $\Phi$ defining the sigma-model can both be taken to be Riemannian. 

The cotangent bundle $T^*\Phi$ can always be locally trivialised, namely  
\begin{equation}
   T^*\Phi\bigg|_{_U}\ \simeq\ U\times\ \mathbf{R}^{^n}.
\end{equation}   

Given $g_{_{ij}}$ on $\Phi$, the Hamiltonian function is defined as follows 

\begin{equation}
H(q,p)\ \overset{def.}{=}\ \frac{1}{2}\ g^{^{ij}}(q)\ p_{_i}p_{_j},
\end{equation} 
from which the Hamilton-Jacobi equations define the geodesic (or Hamiltonian) flow on $\Phi$ 

\begin{equation}   
\dot q^{^i}=\frac{\partial H}{\partial p_{_i}}\ \ \ ,\ \ \ \dot p_{_i}=-\frac{\partial H}{\partial q^{^i}}.  
\end{equation} 

The Hamiltonian description therefore enables us to interpret the sigma-model as the gluing of two energy functionals, namely the momenta in $T^*\Phi$ and $T^*M$.

\subsection*{The sigma-model}  

By definition, \cite{Freed:1999mn}, a sigma-model is a  field theory describing the field as a point particle confined to move on a fixed manifold. The latter could either be a Riemannian manifold, a Lie group, or a symmetric space. 

When coupling it to a gauge field, we get the Landau-Ginzburg (LG)-model. 

The Lagrangian density for the sigma-model can be expressed in many different ways. One of the simplest expressions is the following   

\begin{equation}   
{\cal L}=\frac{1}{2}\sum_{i=1}^{n}\sum_{_j=1}^{n}\ g_{_{ij}}(\phi)\partial^{^{\mu}}\phi_{_i}\partial_{_{\mu}}\phi_{_j}, 
\end{equation}
where $g_{_{ij}}(\phi)$ denotes the metric tensor on the field space $\phi\ \in\ \Phi$, where $\Phi$ can be any Riemannian manifold.

In a more fully-geometric notation, it can be written as a fiber bundle with fibers $\Phi$ over a differentiable manifold $M$. Given a section 

\begin{equation}  
\phi:\ M\ \rightarrow\ \Phi,  
\end{equation}
fix a point $x\ \in\ M$. Then, the pushforward at $x$ is a map of tangent bundles 

\begin{equation}  
d_{_{x}}\phi:\ T_{_{x}}M\ \rightarrow\ T_{_{\phi(x)}}\Phi,  
\end{equation}  
taking 

\begin{equation}  
\partial_{_{\mu}}\ \mapsto\ \frac{\partial \phi^{^i}}{\partial x^{^{\mu}}}\partial _{_i}.  
\end{equation}  

The sigma model action then reduces to   

\begin{equation}  
S=\frac{1}{2}\ \int_{_M}d\phi\ \wedge\ \star\ d\phi. 
\label{eq:clsm}
\end{equation}

Among the most important interpretation of the classical sigma-model is that of non-interacting QM. Taking $\Phi\equiv \mathbf{C}$, \eqref{eq:clsm} becomes 
\begin{equation}  
S=\frac{1}{2}\ <<\phi\ ,\ \Delta\ \phi>>, 
\label{eq:clsm1}
\end{equation}
with

\begin{equation}
   \phi:\ M\rightarrow\mathbb{C} 
\end{equation}
interpreted as a wavefunction, whose Laplacian corresponds to its associated kinetic energy. Hence, the classical sigma-model on $\mathbf{C}$ can be interpreted as the QM of a free particle.

\subsection{Moduli spaces}

In algebraic geometry, a moduli space is a geometric space (a scheme or an algebraic stack) whose points represent algebro-geometric objects of some kind, or their isomorphic classes. The main motivation for introducing moduli spaces is that of finding solutions of geometric problems.

A scheme is a mathematical structure that enlarges the notion of an algebraic variety, accounting for multiplicities, and allowing varieties defined on any commutative ring. Scheme theory was introduced by A. Grothendieck in 1960 for addressing deep problems in algebraic geometry. 

A scheme is a topological space together with commutative rings for all of its open sets, which arises from gluing together spectra of commutative rings along their open subsets, i.e. ringed space which is locally a spectrum of a commutative ring.

Algebraic surfaces can be studied making use of morphisms of schemes. In many cases, the family of all varieties of a given type can itself be viewed as a variety of schemes, known as moduli space.

\subsection*{The nonlinear sigma-model} 

Considering Minkowski spacetime, and denoting with ${\cal F}$ the space of fields, $L$ the Lagrangian, and $\gamma$ the variational 1-fomr, ${\cal M}$ the space of classical solutions, and $\omega$ the local symplectic form. In its classical formulation, the Lagrangian encodes a classical Hamiltonian system once a specified time direction has been selected, thereby breaking Poincare' invariance. 

Upon taking a real-valued scalar field 

\begin{equation}   
\phi: M^{^n}\ \rightarrow\ \mathbb{R}.     
\end{equation}  

The kinetic term in the Lagrangian would therefore read  

\begin{equation} 
\begin{aligned}
L_{_{kin}} &=\frac{1}{2}|d\phi|^{^2}\left|d^{^n}x\right| \\    
&=\frac{1}{2}g^{^{\mu\nu}}\partial_{_{\mu}}\phi\partial_{_{\nu}}\phi \left|d^{^n}x\right| \\  
&=\frac{1}{2}\left[(\partial_{_0}\phi)^{^2}-\sum_{_{i=1}}^{^{n-1}}(\partial_{_i}\phi)^{^2}\right]\left|d^{^n}x\right|.  
\label{eq:lkin}
\end{aligned}   
\end{equation}    

Adding a potential term of degree $\le 2$ of the kind

\begin{equation}  
V: \mathbb{R}\ \longrightarrow\ \mathbb{R}   
\end{equation}  
to \eqref{eq:lkin}, and assuming $V$ is bounded below by choosing suitable coefficients of th e polynomial, then one can eliminate the linear and constant terms by assuming the minimum of $V$ is at the origin with $V=0$ there. The total Lagrangian therefore becomes  

\begin{equation}  
L=\left(\frac{1}{2}|d\phi|^{^2}-\frac{m^{^2}}{2}\phi^{^2}\right)\left|d^{^n}x\right|.  
\end{equation}

In non-free theories, the potential is no longer quadratic. The simplest example of an interacting theory is the case of the quartic potential. In either case, in a pure mathematical formulation, the non-linear $\sigma$-model is characterised by the following data:  

\begin{enumerate}

\item A Riemannian manifold, $X$. 

\item   A potential energy function  

\begin{equation}    
V:\ X\ \longrightarrow\ \mathbb{R}.   
\end{equation}   

\end{enumerate}  

In this formulation, the space of fields is th emapping space    

\begin{equation}  
{\cal F}=\left(\text{Map}(M^{^n}, X\right).  
\end{equation}

The total Lagrangian therefore reads

\begin{equation}  
L=\left[\frac{1}{2}|d\phi|^{^2}-\phi^{^*}V\right]\left|d^{^n}x\right|,      
\end{equation}   
from which the energy density reads

\begin{equation}  
{\cal E}(\phi)=\left[\frac{1}{2}\left|\partial_{_0}\phi\right|^{^2}+\sum_{_{1=1}}^{^{n-1}}\frac{1}{2}\left|\partial_{_i}\phi\right|^{^2}+V\right]\left|d^{^{n-1}}x\right|,      
\end{equation}

The space of solutions, namely the moduli space of vacua, ${\cal M}_{_vac}$, is 

\begin{equation}   
{\cal M}_{_vac}\ \subset\ {\cal F}{\cal E}_{_N},  
\end{equation}
with ${\cal F}{\cal E}_{_N}$ denoting the space of static fields of finite energy. 

For the case of a real scalar field with potential $V$, one usually assumes that the minimum value is at $V=0$, and therefore that the solutions lie at

\begin{equation}  
{\cal M}\ \overset{def.}{=}\ V^{^{-1}}(0).    
\end{equation}  

For example, if $X\equiv\mathbb{R}$, and $V=0$, the resulting theory is that of a massless real scalar field. For this case, one finds that ${\cal M}_{_{vac}}\simeq\mathbb{R}$. On the other hand, for the case of a scalar field with potential 

\begin{equation}   
V(\phi)=\frac{m^{^2}}{2}\phi^{^2}   
\end{equation}  
${\cal M}_{_{vac}}$ is a single point $\phi=0$. For a quartic potential, instead, ${\cal M}_{_{vac}}$ is constituted by two points, i.e. the values of $\phi$ where $V$ is minimised.

\subsection{Kostant-Whittaker reduction}

Let $\mathfrak{g}$ be the Lie algebra of $G$ and fix a regular $\mathfrak{sl}_{_2}$-triple $\{e, \mathfrak{h}, f\}$. The regular nilpotent element $e$ is contained in a unique Borel subalgebra $\mathfrak{b}$, with corresponding subgroup $U$. 

Assuming we start from a Poisson manifold $M$ with Hamiltonian $G$-action, the associated moment map reads 

\begin{equation}  
\mu:\ M\ \rightarrow\ \mathfrak{g}  
\end{equation}  

The Whittaker reduction of $M$ is the Poisson manifold  

\begin{equation}   
\mu^{^{-1}}(\mathfrak{f+b})\bigg/U  
\end{equation}  
obtained by Hamiltonian reduction with respect to the action of $U$ at a point corresponding to the regular character.  

On the other hand, the principal slice of regular elements is 

\begin{equation}  
S\overset{def.}{=}\mathfrak{f}+\mathfrak{g}^{^e}\ \subset\ \mathfrak{g}  
\end{equation}  
as defined by Kostant. By definition, $S$ is contained in the space $\mathfrak{f}+\mathfrak{b}$, therefore we obtain 

\begin{equation}  
\mu^{^{-1}}(S)\ \simeq\ \mu^{^{-1}}(\mathfrak{f+b})\ \bigg/\ U  
\end{equation}   
leading to a canonical relative compactification of the universal centraliser $Z_{_G}$, resulting in a symplectic variety obtained by Whittaker reduction relative to the $G\times G$-action on the cotangent bundle $T^{^*}G_{_{ad}}$.

\subsection{Moore-Tachikawa varieties}   \label{sec:mt}

This subsection is structured as follows:   

\begin{enumerate}

\item  At first, we briefly overview the source and target categorical structure proposed in \cite{Moore:2011ee} assuming duality.   

\item  We then explain what categorical duality means from an algebraic perspective.

\item  We conclude the section indicating the relation between Moore-Tachikawa varieties and Coulomb branches of quiver gauge theories as an interesting realisation of 3D mirror symmetry, \cite{Intriligator:1996ex}, and how the categorical generalisation proposed in this work suggests interesting applications to quiver varieties that will be addressed in more detail in Part III.

\end{enumerate}

\subsection*{Categorical structure assuming duality}  

As already mentioned in the introduction to Part V, to a given class ${\cal S}$ theory, one can assign a 2D TFT valued in a symmetric monoidal category, \cite{Moore:2011ee},

\begin{equation}   
\boxed{\ \ \ \eta_{_{G_{\mathbb{C}}}}:\ \text{Bo}_{_2}\ \rightarrow\ \text{HS} \color{white}\bigg]\ \ \ }   
\label{eq:etaGC}
\end{equation}

The existence of this 2D TFT relies on the the source and target categories satisfying a certain list of properties, \cite{Moore:2011ee}. We will not reproduce all of them in our treatment, and refer the interested reader to the original work of Moore and Tachikawa for a detailed explanation. In this first part of the section, we will only point out some of the crucial assumptions made in their work for reasons that will become clear in the following pages. 

\section*{Duality}

\begin{figure}[ht!]       
\begin{center}  
\includegraphics[scale=1]{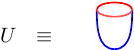}   \ \ \  \ \ \ 
  \includegraphics[scale=1]{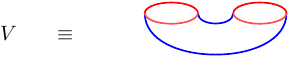} \ \ \   \ \ \ 
\includegraphics[scale=1]{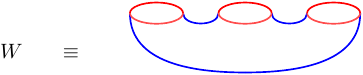}   
\caption{\small Basic bordisms assuming duality of, both, the source and target categories leading to the definition of the identity element, $V_{_{G_{_{\mathbb{C}}}}}$, and the maximal dimensional Higgs branch, $W_{_{G_{_{\mathbb{C}}}}}$.}    
\label{figure:UW}    
\end{center}  
\end{figure}

For the purpose of our work, the crucial assumption made in \cite{Moore:2011ee} is the duality structure of the source category Bo$_{_2}$. As explained in \cite{Moore:2011ee}, duality implies that the 2-category Bo$_{_2}$ is fully specified by its objects, $S^{1}$, and 1-morphisms, namely the bordisms depicted in figure \ref{figure:UW}. The middle bordism, i.e. the one labelled $V$, is the identity bordism. One can easily see this by noticing that $V$ is topologically equivalent to a cylinder whose edges are the red circles, i.e. the object of 2-category Bo$_{_2}$.

For $\eta_{_{G_{\mathbb{C}}}}$ to be well defined, the source and target categories are required to satisfy certain sewing relations, \cite{Moore:2006dw,Moore:2011ee}. This practically means that, compositions between morphisms should close. In particular, the identity itself can be defined in terms of composite homomorphisms as follows,

\begin{equation}             
\boxed{\ \ \ U_{_{G_{\mathbb{C}}}}\ \circ\ W_{_{G_{\mathbb{C}}}}\ \equiv\ T^{^*}G_{_{\mathbb{C}}} \color{white}\bigg]\ },  
\label{eq:defofid}     
\end{equation} 
where $T^{^*}G_{_{\mathbb{C}}}\ \equiv\ V{_{G_{\mathbb{C}}}}$.
Indeed, one can easily see that combining the first and third bordisms in figure \ref{figure:UW}, is topologically equivalent to $V$.\footnote{Indeed, $V$ is topologically equivalent to the cylinder, i.e. the cobordism between $S^{^1}$ and itself.} 

\subsection*{Key axiom}  

\eqref{eq:defofid} is essential for us in relating the formalism of \cite{Moore:2006dw,Moore:2011ee} to the setup of figure \ref{fig:correspondence}. In particular, it is what leads to the definition of the triple featuring on the RHS of \eqref{eq:HB1}. To see this explicitly, let us recall a crucial axiom required to be satisfied by \eqref{eq:etaGC}, and, therefore, in turn by \eqref{eq:defofid}, \cite{Moore:2011ee}.

For $X\in$\ Hom\ $(G_{_{\mathbb{C}}}^{^{\prime}}, G_{_{\mathbb{C}}})$ and $Y\in$\ Hom\ $(G_{_{\mathbb{C}}}, G_{_{\mathbb{C}}}^{^{\prime\prime}})$, their composition 

\begin{equation}   
Y\ \circ\ X\ \in\ \text{Hom}\ (G_{_{\mathbb{C}}}^{^{\prime}}, G_{_{\mathbb{C}}}^{^{\prime\prime}}) 
\label{eq:etaGC1}
\end{equation}    
is identified with the holomorphic symplectic quotient   

\begin{equation}   
\begin{aligned}
Y\ \circ\ X\ &\overset{def.}{=}\ X\ \times\ Y\ //\ G_{_{\mathbb{C}}}   \\
&\ =\{(x,y)\ \in\ X\times Y|\ \mu_{_X}(x)+\mu_{_Y}(y)=0\}\ /\ G_{_{\mathbb{C}}},     
\end{aligned}
\label{eq:etaGC1}
\end{equation}  
where

\begin{equation}   
\mu_{_X}:\ X\ \longrightarrow\ \mathfrak{g}_{_{\mathbb{C}}}^{*}\ \ \ , \ \ \ \mu_{_Y}:\ Y\ \longrightarrow\ \mathfrak{g}_{_{\mathbb{C}}}^{*}
\label{eq:etaGC1b}
\end{equation}  
are the moment maps of the action of $G_{_{\mathbb{C}}}$ on $X$ and $Y$, with $\mathfrak{g}_{_{\mathbb{C}}}$ the Lie algebra associated to $G_{_{\mathbb{C}}}$.  The identity element

\begin{equation}   
T^{^*}G_{_{\mathbb{C}}}\ \overset{def.}{=}\ \text{id}_{_{G_{_{\mathbb{C}}}}}\ \in\ \text{Hom}\ (G_{_{\mathbb{C}}}, G_{_{\mathbb{C}}}) 
\label{eq:etaGC2}
\end{equation}           
comes with a Hamiltonian $G_{_{\mathbb{C}}}\ \times\ G_{_{\mathbb{C}}}$ action. As also explained in \cite{Moore:2011ee}, to see that $T^{^*}G_{_{\mathbb{C}}}$ acts as the identity, it is enough to consider a composition of homomorphisms, $T^{^*}G_{_{\mathbb{C}}}\ \circ\ X$. Identifying $T^{^*}G_{_{\mathbb{C}}}\ \simeq\ G_{_{\mathbb{C}}}\ \times\ \mathfrak{g}_{_{\mathbb{C}}}$, and an element of $T^{^*}G_{_{\mathbb{C}}}$ as $(g,a)$.   The moment map condition, \eqref{eq:etaGC1}, reduces to   

\begin{equation}   
a+\mu(x)=0,      
\label{eq:etaGC3}
\end{equation}  
from which $a$ can be removed. Consequently, the induced 2-form on the solution space is $G_{_{\mathbb{C}}}$-invariant and basic. Upon taking the quotient with respect to $G_{_{\mathbb{C}}}$, we can gauge $g$ to 1, leading to a holomorphic isomorphism with the original $X$ space with its symplectic form.

The categorical quotient taken in defining the composition \eqref{eq:etaGC1} is equivalent to the one that a given absolute theory should be equipped with to potentially gauge away its entire operator content, while leaving only the identity in the spectrum\footnote{cf. explanation in the Introduction. This is basically what leads to the definition of the fiber functor, moment map and Drinfeld center.}. Indeed, this is true as long as the embeddings of the subalgebras associated to $G_{_{\mathbb{C}}}^{^{\prime}}$ and $G_{_{\mathbb{C}}}^{^{\prime\prime}}$ are subsets of each other within the mother algebra $\mathfrak{g}_{_{\mathbb{C}}}$. However, we are interested in describing more general setups, where the embeddings of the algebras are intersecting albeit not one included within the other. In the remainder of our treatment, we will explain that, for this to be described in the formalism of \cite{Moore:2006dw,Moore:2011ee}, the standard identity element associated to the gauge group $G_{_{\mathbb{C}}}$ and embedding Lie algebra $g_{_{\mathbb{C}}}$ needs to be removed from Bo$_{_2}$, while being replaced by a new composite bordism, and propose the definition of a new functor.

\subsection*{Algebraic Varieties} \label{sec:FBFN}

In the concluding part of this section, we highlight an interesting relation to 3D mirror symmetry and generalisations thereof, presented by the same author in \cite{Pasquarella:2023exd}.

The Higgs branches described by Moore and Tachikawa are known to have been reproduced by \cite{Braverman:2017ofm} as the Coulomb branches of 3D ${\cal N}=4$ supersymmetric quiver gauge theories. Such correspondence is therefore equivalent to a statement of 3D mirror symmetry. The purpose of \cite{Pasquarella:2023} is to explain what the 3D dual of a theory obtained relaxing the categorical duality assumption is in terms of Coulomb branches of 3D ${\cal N}=4$ quiver gauge theories. This enabled us to prove the statements made in \cite{Pasquarella:2023exd}.

If $V$ has been removed from the source, one should expect there to be more than one 2D TFT 
 of the kind \eqref{eq:etaGC} associated to two different gauge groups whose embedding in the gauge group associated to the original TFT with identity element $V$ is not simply a cochain complex. Correspondingly, this also means that there is more than one 1-morphism $U_{_{G_{\mathbb{C}}}}$.

Given that the identity of the embedding theory is defined as follows

\begin{equation}             
\boxed{\ \ \ U_{_{G_{\mathbb{C}}}}\ \circ\ W_{_{G_{\mathbb{C}}}}\ \equiv\ T^{^*}G_{_{\mathbb{C}}} \color{white}\bigg]\ }\ \ \ \ , \ \ \ \  
\boxed{\ \ \ \eta_{_{G_{\mathbb{C}}}}(V)\ \equiv\ T^{^*}G_{_{\mathbb{C}}}\color{white}\bigg]\ \ }.    
\end{equation} 
and that 

\begin{equation}             
U_{_{G_{\mathbb{C}}}}\ \overset{def.}{=}\ G_{_{\mathbb{C}}}\times S_{_{n}}\ \subset\ G_{_{\mathbb{C}}}\ \times\ \mathfrak{g}_{_{\mathbb{C}}}\ \simeq\ T^{^*}G_{_{\mathbb{C}}},    
\end{equation} 
with $S_{_n}$ is the Slodowy slice at a principal nilpotent element $n$. The physical theories of class ${\cal S}$ predict the existence of a variety $W_{_{G_{_{\mathbb{C}}}}}$ satisfying the properties needed to define a TFT, $\eta_{_{G_{\mathbb{C}}}}$.  From the duality assumption, it follows that the dimensionalities of the two varieties are related as follows

\begin{equation}             
\text{dim}_{_{\mathbb{C}}}\ U_{_{G_{\mathbb{C}}}}\ \overset{def.}{=}\ \text{dim}_{_{\mathbb{C}}}\ G_{_{\mathbb{C}}}\ +\ \text{rank}\ G_{_{\mathbb{C}}}.  
\end{equation} 

\begin{equation}             
\text{dim}_{_{\mathbb{C}}}\ W_{_{G_{\mathbb{C}}}}\ \overset{def.}{=}\ 3\ \text{dim}_{_{\mathbb{C}}}\ G_{_{\mathbb{C}}}\ -\ \text{rank}\ G_{_{\mathbb{C}}}. 
\label{eq:dimhb}   
\end{equation} 

However, if the identity needs to be removed from Bo$_{_2}$, $T^{^*}G_{_{\mathbb{C}}}$ is not the identity and, in particular \eqref{eq:dimhb} needs to be redefined precisely because the source is no longer a dual category.

As a concluding remark to what we have just said, in \cite{Moore:2006dw} they conjecture the following property for the moment maps associated to the $G^{^3}$ action on the Higgs branch $W_{_{G_{_{\mathbb{C}}}}}$

\begin{equation}   
\boxed{\ \ \ \mu_{_i}:\ W_{_{G_{_{\mathbb{C}}}}}\rightarrow \mathfrak{g}_{_{\mathbf{C}}}^{^*}\ \ \ ,\ \ \ i=1,2,3.  \color{white}\bigg]\ \ }  
\label{eq:HBB}  
\end{equation}   

This is crucial to our analysis since \eqref{eq:HBB} can be inverted to obtain the Higgs branch as a hyperk$\ddot{\text{a}}$ler quotient 

\begin{equation}   
\boxed{\ \ \ W_{_{G_{_{\mathbb{C}}}}}\ \equiv \ \mu^{^-1} /G^{^3}\color{white}\bigg]\ \ }.  
\label{eq:HBtrue}  
\end{equation}

However, for the case in which the identity is removed from the source category, \eqref{eq:HBtrue}  does not hold anymore precisely because of the lack of permutational symmetry arising in the quotient. In \cite{Pasquarella:2023} we explained how to define $W_{_{G_{_{\mathbb{C}}}}}$ and its dimensionality for the case involving categories without a duality structure by making use of factorisation homology.


\begin{thebibliography}{99} 




\bibitem{Nilles:1995ci}
H.~P.~Nilles,
\emph{Phenomenological aspects of supersymmetry},
[arXiv:hep-ph/9511313 [hep-ph]].  

\bibitem{Csaki:1996ks}
C.~Csaki,
\emph{The Minimal supersymmetric standard model (MSSM)},
Mod. Phys. Lett. A \textbf{11} (1996), 599
doi:10.1142/S021773239600062X
[arXiv:hep-ph/9606414 [hep-ph]].



\bibitem{Slavich:2020zjv}
P.~Slavich, S.~Heinemeyer, E.~Bagnaschi, H.~Bahl, M.~Goodsell, H.~E.~Haber, T.~Hahn, R.~Harlander, W.~Hollik and G.~Lee, \textit{et al.}
\emph{Higgs-mass predictions in the MSSM and beyond},
Eur. Phys. J. C \textbf{81} (2021) no.5, 450
doi:10.1140/epjc/s10052-021-09198-2
[arXiv:2012.15629 [hep-ph]].  

\bibitem{LHCReinterpretationForum:2020xtr}
W.~Abdallah \textit{et al.} [LHC Reinterpretation Forum],
\emph{Reinterpretation of LHC Results for New Physics: Status and Recommendations after Run 2}, 
SciPost Phys. \textbf{9} (2020) no.2, 022
doi:10.21468/SciPostPhys.9.2.022
[arXiv:2003.07868 [hep-ph]].


\bibitem{Allanach:2021bbd}
B.~C.~Allanach and H.~Banks,
\emph{Hide and seek with the third family hypercharge model\textquoteright{}s $Z^\prime $ at the large hadron collider},
Eur. Phys. J. C \textbf{82} (2022) no.3, 279
doi:10.1140/epjc/s10052-022-10191-6
[arXiv:2111.06691 [hep-ph]].


\bibitem{Banks:2020gpu}
\emph{Charting the Fifth Force Landscape},
Phys. Rev. D \textbf{103} (2021) no.7, 075018
doi:10.1103/PhysRevD.103.075018
[arXiv:2009.12399 [hep-ph]].

\bibitem{Allanach:2023bgg}
B.~Allanach,
\emph{Fits to measurements of rare heavy flavour decays},
[arXiv:2307.07532 [hep-ph]].

\bibitem{Hammou:2023heg}
E.~Hammou, Z.~Kassabov, M.~Madigan, M.~L.~Mangano, L.~Mantani, J.~Moore, M.~M.~Alvarado and M.~Ubiali,
\emph{Hide and seek: how PDFs can conceal new physics},
JHEP \textbf{11} (2023), 090
doi:10.1007/JHEP11(2023)090
[arXiv:2307.10370 [hep-ph]].


\bibitem{Kassabov:2023hbm}
Z.~Kassabov, M.~Madigan, L.~Mantani, J.~Moore, M.~Morales Alvarado, J.~Rojo and M.~Ubiali,
\emph{The top quark legacy of the LHC Run II for PDF and SMEFT analyses},
JHEP \textbf{05} (2023), 205
doi:10.1007/JHEP05(2023)205
[arXiv:2303.06159 [hep-ph]].


\bibitem{Iranipour:2022iak}
S.~Iranipour and M.~Ubiali,
\emph{A new generation of simultaneous fits to LHC data using deep learning},
JHEP \textbf{05} (2022), 032
doi:10.1007/JHEP05(2022)032
[arXiv:2201.07240 [hep-ph]].


\bibitem{Nath:2010zj}
P.~Nath, B.~D.~Nelson, H.~Davoudiasl, B.~Dutta, D.~Feldman, Z.~Liu, T.~Han, P.~Langacker, R.~Mohapatra and J.~Valle, \textit{et al.}
\emph{The Hunt for New Physics at the Large Hadron Collider},
Nucl. Phys. B Proc. Suppl. \textbf{200-202} (2010), 185-417
doi:10.1016/j.nuclphysbps.2010.03.001
[arXiv:1001.2693 [hep-ph]].   


\bibitem{Allanach:2005pv}
B.~C.~Allanach, F.~Quevedo and K.~Suruliz,
\emph{Low-energy supersymmetry breaking from string flux compactifications: Benchmark scenarios}, 
JHEP \textbf{04} (2006), 040
doi:10.1088/1126-6708/2006/04/040
[arXiv:hep-ph/0512081 [hep-ph]].







\bibitem{Cicoli:2021dhg}
M.~Cicoli, I.~G.~Etxebarria, F.~Quevedo, A.~Schachner, P.~Shukla and R.~Valandro,
\emph{The Standard Model quiver in de Sitter string compactifications},
JHEP \textbf{08} (2021), 109
doi:10.1007/JHEP08(2021)109
[arXiv:2106.11964 [hep-th]].




\bibitem{Krippendorf:2010hj}
S.~Krippendorf, M.~J.~Dolan, A.~Maharana and F.~Quevedo,
\emph{D-branes at Toric Singularities: Model Building, Yukawa Couplings and Flavour Physics},
JHEP \textbf{06} (2010), 092
doi:10.1007/JHEP06(2010)092
[arXiv:1002.1790 [hep-th]].








\bibitem{AbdusSalam:2009qd}
S.~S.~AbdusSalam, B.~C.~Allanach, F.~Quevedo, F.~Feroz and M.~Hobson,
\emph{Fitting the Phenomenological MSSM},
Phys. Rev. D \textbf{81} (2010), 095012
doi:10.1103/PhysRevD.81.095012
[arXiv:0904.2548 [hep-ph]].


\bibitem{AbdusSalam:2007pm}
S.~S.~AbdusSalam, J.~P.~Conlon, F.~Quevedo and K.~Suruliz,
\emph{Scanning the Landscape of Flux Compactifications: Vacuum Structure and Soft Supersymmetry Breaking,},
JHEP \textbf{12} (2007), 036
doi:10.1088/1126-6708/2007/12/036
[arXiv:0709.0221 [hep-th]].




\bibitem{Conlon:2007xv}
J.~P.~Conlon, C.~H.~Kom, K.~Suruliz, B.~C.~Allanach and F.~Quevedo,
\emph{Sparticle Spectra and LHC Signatures for Large Volume String Compactifications},
JHEP \textbf{08} (2007), 061
doi:10.1088/1126-6708/2007/08/061
[arXiv:0704.3403 [hep-ph]].

\bibitem{Cremades:2007ig}
D.~Cremades, M.~P.~Garcia del Moral, F.~Quevedo and K.~Suruliz,
\emph{Moduli stabilisation and de Sitter string vacua from magnetised D7 branes},
JHEP \textbf{05} (2007), 100
doi:10.1088/1126-6708/2007/05/100
[arXiv:hep-th/0701154 [hep-th]].










\bibitem{gsw1}
Green MB, Schwarz JH, Witten E. Superstring Theory: 25th Anniversary Edition. Cambridge University Press; 2012. Vol. 1

\bibitem{gsw2}
Green MB, Schwarz JH, Witten E. Superstring Theory: 25th Anniversary Edition. Cambridge University Press; 2012. Vol.2

\bibitem{jp1}
Polchinski J. String Theory. Cambridge University Press; 1998. Vol.1

\bibitem{jp2}
Polchinski J. String Theory. Cambridge University Press; 1998. Vol. 2    


\bibitem{Polchinski:1994mb}
J.~Polchinski,
\emph{What is string theory?}, 
[arXiv:hep-th/9411028 [hep-th]].



\bibitem{Green:1982ct}
M.~B.~Green,
\emph{Supersymmetrical Dual String Theories and their Field Theory Limits: A Review}, 
Surveys High Energ. Phys. \textbf{3} (1984), 127-160
doi:10.1080/01422418308243456

\bibitem{Schwarz:1982jn}
J.~H.~Schwarz,
\emph{Superstring Theory}, 
Phys. Rept. \textbf{89} (1982), 223-322
doi:10.1016/0370-1573(82)90087-4


\bibitem{Green:1984sg}
M.~B.~Green and J.~H.~Schwarz,
\emph{Anomaly Cancellation in Supersymmetric D=10 Gauge Theory and Superstring Theory}, 
Phys. Lett. B \textbf{149} (1984), 117-122
doi:10.1016/0370-2693(84)91565-X

\bibitem{Gross:1984dd}
D.~J.~Gross, J.~A.~Harvey, E.~J.~Martinec and R.~Rohm,
\emph{The Heterotic String}, 
Phys. Rev. Lett. \textbf{54} (1985), 502-505
doi:10.1103/PhysRevLett.54.502   



\bibitem{VP}    
V. Pasquarella, to appear.  



\bibitem{Braverman:2017ofm}
A.~Braverman, M.~Finkelberg and H.~Nakajima,
\emph{Ring objects in the equivariant derived Satake category arising from Coulomb branches (with an appendix by Gus Lonergan)}, 
[arXiv:1706.02112 [math.RT]].

\bibitem{Braverman:2016wma}
A.~Braverman, M.~Finkelberg and H.~Nakajima,
\emph{Towards a mathematical definition of Coulomb branches of $3$-dimensional $\mathcal{N} = 4$ gauge theories, II},    
Adv. Theor. Math. Phys. \textbf{22} (2018), 1071-1147
doi:10.4310/ATMP.2018.v22.n5.a1
[arXiv:1601.03586 [math.RT]].





\bibitem{Freed:2022qnc}
D.~S.~Freed, G.~W.~Moore and C.~Teleman,
\emph{Topological symmetry in quantum field theory},   
[arXiv:2209.07471 [hep-th]]. 

\bibitem{Teleman:2014jaa}
C.~Teleman,
\emph{Gauge theory and mirror symmetry},
[arXiv:1404.6305 [math-ph]].

\bibitem{Pasquarella:2023exd}
V.~Pasquarella,
\emph{Drinfeld Centers from Magnetic Quivers},
[arXiv:2306.12471 [hep-th]].


\bibitem{Pasquarella:2023}
V.~Pasquarella,
\emph{Factorisation Homology of Class ${\cal S}$ Theories},
[arXiv:2312. [hep-th]].

\bibitem{Moore:2011ee}
G.~W.~Moore and Y.~Tachikawa,
\emph{On 2d TQFTs whose values are holomorphic symplectic varieties},    
Proc. Symp. Pure Math. \textbf{85} (2012), 191-208
doi:10.1090/pspum/085/1379
[arXiv:1106.5698 [hep-th]].    


\bibitem{Webster:2016rhh}
\emph{Koszul duality between Higgs and Coulomb categories $\mathcal{O}$},   
[arXiv:1611.06541 [math.RT]].




\bibitem{Pasquarella:2023ntw}
V.~Pasquarella,
\emph{Moore-Tachikawa Varieties: Beyond Duality},   
JHAP \textbf{3} (2023) no.4, 39-56
doi:10.22128/jhap.2023.741.1061
[arXiv:2310.01489 [hep-th]].





\bibitem{W1}
Weinberg S. The Quantum Theory of Fields. Cambridge University Press; 1995. Vol. 1


\bibitem{W2}
Weinberg S. The Quantum Theory of Fields. Cambridge University Press; 1996. Vol. 2

\bibitem{W3}
Weinberg S. The Quantum Theory of Fields. Cambridge University Press; 2000. Vol. 3


\bibitem{Weinberg:2004kv}
S.~Weinberg,
\emph{The Making of the standard model},
Eur. Phys. J. C \textbf{34} (2004), 5-13
doi:10.1140/epjc/s2004-01761-1
[arXiv:hep-ph/0401010 [hep-ph]].



\bibitem{Quevedo:2010ui}
F.~Quevedo, S.~Krippendorf and O.~Schlotterer,
\emph{Cambridge Lectures on Supersymmetry and Extra Dimensions},
[arXiv:1011.1491 [hep-th]].





 
\bibitem{Freed:2012bs}
D.~S.~Freed and C.~Teleman,
\emph{Relative quantum field theory},    
Commun. Math. Phys. \textbf{326} (2014), 459-476
doi:10.1007/s00220-013-1880-1
[arXiv:1212.1692 [hep-th]].   

\bibitem{Dimofte:2018abu}
T.~Dimofte and N.~Garner,
\emph{Coulomb Branches of Star-Shaped Quivers},
JHEP \textbf{02} (2019), 004
doi:10.1007/JHEP02(2019)004
[arXiv:1808.05226 [hep-th]].  


\bibitem{Bullimore:2015lsa}
M.~Bullimore, T.~Dimofte and D.~Gaiotto,
\emph{The Coulomb Branch of 3d ${\mathcal{N}= 4}$ Theories},   
Commun. Math. Phys. \textbf{354} (2017) no.2, 671-751
doi:10.1007/s00220-017-2903-0
[arXiv:1503.04817 [hep-th]].


\bibitem{Jordan:2023gtq}
D.~Jordan,
\emph{Quantum character varieties},
[arXiv:2309.06543 [math.QA]].






\bibitem{Graf:2022rco}
L.~Gr\'af, B.~Henning, X.~Lu, T.~Melia and H.~Murayama,
\emph{Hilbert series, the Higgs mechanism, and HEFT},  
JHEP \textbf{02} (2023), 064
doi:10.1007/JHEP02(2023)064
[arXiv:2211.06275 [hep-ph]].


\bibitem{Bento:2023owf}
M.~P.~Bento, J.~P.~Silva and A.~Trautner,
\emph{The basis invariant flavor puzzle},  
JHEP \textbf{01} (2024), 024
doi:10.1007/JHEP01(2024)024
[arXiv:2308.00019 [hep-ph]].

\bibitem{Anisha:2019nzx}
Anisha, S.~Das Bakshi, J.~Chakrabortty and S.~Prakash,
\emph{Hilbert Series and Plethystics: Paving the path towards 2HDM- and MLRSM-EFT},  
JHEP \textbf{09} (2019), 035
doi:10.1007/JHEP09(2019)035
[arXiv:1905.11047 [hep-ph]].


\bibitem{Freed:2006tm}
D.~S.~Freed,
\emph{Classical field theory and supersymmetry}.




\bibitem{Higgs:1964ia}
P.~W.~Higgs,
\emph{Broken symmetries, massless particles and gauge fields},
Phys. Lett. \textbf{12} (1964), 132-133
doi:10.1016/0031-9163(64)91136-9


\bibitem{Higgs:1966ev}
P.~W.~Higgs,
\emph{Spontaneous Symmetry Breakdown without Massless Bosons},
Phys. Rev. \textbf{145} (1966), 1156-1163
doi:10.1103/PhysRev.145.1156








 \bibitem{K.Wilson} 
 K. Wilson, Rev. Mod. Phys. 55, 583 (1983).






\bibitem{TJF}    
T.~Johnson-Freyd,    
\emph{Operators and higher categories in quantum field theory}, Lecture series.   






\bibitem{Uranga}   
Conference Proceedings
 From Quiver Diagrams to Particle Physics
Uranga, Angel M.
 Casacuberta, Carles
 Miró-Roig, Rosa Maria
 Verdera, Joan
 Xambó-Descamps, Sebastià
European Congress of Mathematics
2001
 Birkhäuser Basel
 Basel
978-3-0348-8266-8
 10.1007/978-3-0348-8266-8 43

\bibitem{Allanach:2021bfe}
B.~C.~Allanach, B.~Gripaios and J.~Tooby-Smith,
\emph{Semisimple extensions of the Standard Model gauge algebra},
Phys. Rev. D \textbf{104} (2021) no.3, 035035
[erratum: Phys. Rev. D \textbf{106} (2022) no.1, 019901]
doi:10.1103/PhysRevD.104.035035
[arXiv:2104.14555 [hep-th]].




\bibitem{Freed:2019sco}
D.~S.~Freed and M.~J.~Hopkins,
\emph{Consistency of M-Theory on Non-Orientable Manifolds},
Quart. J. Math. Oxford Ser. \textbf{72} (2021) no.1-2, 603-671
doi:10.1093/qmath/haab007
[arXiv:1908.09916 [hep-th]].

\bibitem{Freed:2019jzd}
D.~S.~Freed and M.~J.~Hopkins,
\emph{Invertible phases of matter with spatial symmetry},
Adv. Theor. Math. Phys. \textbf{24} (2020) no.7, 1773-1788
doi:10.4310/ATMP.2020.v24.n7.a3
[arXiv:1901.06419 [math-ph]].

\bibitem{Freed:2018cec}
D.~S.~Freed and C.~Teleman,
\emph{Topological dualities in the Ising model},
Geom. Topol. \textbf{26} (2022), 1907-1984
doi:10.2140/gt.2022.26.1907
[arXiv:1806.00008 [math.AT]].

\bibitem{Freed:2014iua}
D.~S.~Freed,
\emph{Anomalies and Invertible Field Theories},
Proc. Symp. Pure Math. \textbf{88} (2014), 25-46
doi:10.1090/pspum/088/01462
[arXiv:1404.7224 [hep-th]].

\bibitem{Freed:2012hx}
D.~S.~Freed,
\emph{The cobordism hypothesis},
[arXiv:1210.5100 [math.AT]].

\bibitem{Freed:2010wsz}
D.~S.~Freed, M.~J.~Hopkins and C.~Teleman,
\emph{Consistent Orientation of Moduli Spaces},
doi:10.1093/acprof:oso/9780199534920.003.0019





\bibitem{Freed:2009qp}
D.~S.~Freed, M.~J.~Hopkins, J.~Lurie and C.~Teleman,
\emph{Topological Quantum Field Theories from Compact Lie Groups},
[arXiv:0905.0731 [math.AT]].

\bibitem{Freed:2008jq}
D.~S.~Freed,
\emph{Remarks on Chern-Simons Theory},
[arXiv:0808.2507 [math.AT]].



\bibitem{Freed:2006ph}
D.~S.~Freed, D.~R.~Morrison and I.~M.~Singer,
\emph{Quantum field theory, supersymmetry, and enumerative geometry},

\bibitem{Freed:2005qu}
D.~S.~Freed, M.~J.~Hopkins and C.~Teleman,
\emph{Loop groups and twisted K-theory. II.},
[arXiv:math/0511232 [math.AT]].

\bibitem{Freed:2003qx}
D.~S.~Freed, M.~J.~Hopkins and C.~Teleman,
\emph{Twisted K-theory and loop group representations. 1.},
[arXiv:math/0312155 [math.AT]].

\bibitem{Freed:2002qp}
D.~S.~Freed,
\emph{K theory in quantum field theory},
[arXiv:math-ph/0206031 [math-ph]].

\bibitem{Freed:2001jd}
D.~S.~Freed,
\emph{The Verlinde algebra is twisted equivariant K theory},
Turk. J. Math. \textbf{25} (2001), 159-167
[arXiv:math/0101038 [math.RT]].

\bibitem{Freed:2000ta}
D.~S.~Freed,
\emph{Dirac charge quantization and generalized differential cohomology},
[arXiv:hep-th/0011220 [hep-th]].





\bibitem{Freed:2000tt}
D.~S.~Freed and M.~J.~Hopkins,
\emph{On Ramond-Ramond fields and K theory},
JHEP \textbf{05} (2000), 044
doi:10.1088/1126-6708/2000/05/044
[arXiv:hep-th/0002027 [hep-th]].

\bibitem{Freed:1999vc}
D.~S.~Freed and E.~Witten,
\emph{Anomalies in string theory with D-branes},
Asian J. Math. \textbf{3} (1999), 819
[arXiv:hep-th/9907189 [hep-th]].

\bibitem{Deligne:1999qp}
P.~Deligne, P.~Etingof, D.~S.~Freed, L.~C.~Jeffrey, D.~Kazhdan, J.~W.~Morgan, D.~R.~Morrison and E.~Witten,
\emph{Quantum fields and strings: A course for mathematicians. Vol. 1, 2}.

\bibitem{Freed:1999mn}
D.~S.~Freed,
\emph{Five lectures on supersymmetry},
AMS, 1999,
ISBN 978-0-8218-1953-1

\bibitem{Freed:1995ku}
D.~Freed and K.~Uhlenbeck,
\emph{Geometry and quantum field theory. Proceedings, Graduate Summer School on the Geometry and Topology of Manifolds and Quantum Field Theory, Park City, USA, June 22-July 20, 1991}.

\bibitem{Freed:1994ad}
D.~S.~Freed,
\emph{Higher algebraic structures and quantization},
Commun. Math. Phys. \textbf{159} (1994), 343-398
doi:10.1007/BF02102643
[arXiv:hep-th/9212115 [hep-th]].

\bibitem{Freed:1993wb}
D.~S.~Freed,
\emph{Extended structures in topological quantum field theory},
[arXiv:hep-th/9306045 [hep-th]].

\bibitem{Freed:1993in}
D.~S.~Freed,
\emph{Characteristic numbers and generalized path integrals},

\bibitem{Freed:1992qb}
D.~S.~Freed,
\emph{Locality and integration in topological field theory},
[arXiv:hep-th/9209048 [hep-th]].

\bibitem{Freed:1992vw}
D.~S.~Freed,
\emph{Classical Chern-Simons theory. Part 1},
Adv. Math. \textbf{113} (1995), 237-303
doi:10.1006/aima.1995.1039
[arXiv:hep-th/9206021 [hep-th]].


\bibitem{Freed:1992vf}
D.~S.~Freed,
\emph{Lectures on topological quantum field theory},
NATO Sci. Ser. C \textbf{409} (1993), 95-156

\bibitem{Freed:1991bn}
D.~S.~Freed and F.~Quinn,
\emph{Chern-Simons theory with finite gauge group},
Commun. Math. Phys. \textbf{156} (1993), 435-472
doi:10.1007/BF02096860
[arXiv:hep-th/9111004 [hep-th]].

\bibitem{Freed:1984xe}
D.~S.~Freed and K.~K.~Uhlenbeck,
    \emph{INSTANTONS AND FOUR - MANIFOLDS}

\bibitem{Freed:2006ya}
D.~S.~Freed, G.~W.~Moore and G.~Segal,
\emph{The Uncertainty of Fluxes},
Commun. Math. Phys. \textbf{271} (2007), 247-274
doi:10.1007/s00220-006-0181-3
[arXiv:hep-th/0605198 [hep-th]].

\bibitem{Segal:2002ei}
G.~Segal,
\emph{The definition of conformal field theory},

\bibitem{Hitchin:1999at}
N.~J.~Hitchin, G.~B.~Segal and R.~S.~Ward,
\emph{Integrable systems: Twistors, loop groups, and Riemann surfaces. Proceedings, Conference, Oxford, UK, September 1997},

\bibitem{Segal:1996ku}
G.~Segal,
\emph{Space from the point of view of loop groups},

\bibitem{Pressley:1988qk}
A.~Pressley and G.~Segal,
\emph{LOOP GROUPS}

\bibitem{Segal:1987sk}
G.~B.~Segal,
\emph{THE DEFINITION OF CONFORMAL FIELD THEORY}










\bibitem{Lurie:2009keu}
J.~Lurie,
\emph{On the Classification of Topological Field Theories},
[arXiv:0905.0465 [math.CT]]. 

\bibitem{Witten:1995zh}
E.~Witten,
\emph{Some comments on string dynamics},  
[arXiv:hep-th/9507121 [hep-th]].     





\bibitem{Witten:2007ct}
E.~Witten,
\emph{Conformal Field Theory In Four And Six Dimensions},   
[arXiv:0712.0157 [math.RT]].



\bibitem{Deligne-Mumford}
Deligne, Pierre; Mumford, David. The irreducibility of the space of curves of given genus. Publications Mathématiques de l'IHÉS, Volume 36 (1969), pp. 75-109. http://www.numdam.org/item/PMIHES 1969 36 75 0/.











\bibitem{Intriligator:1996ex}
K.~A.~Intriligator and N.~Seiberg,
\emph{Mirror symmetry in three-dimensional gauge theories},
Phys. Lett. B \textbf{387} (1996), 513-519
doi:10.1016/0370-2693(96)01088-X
[arXiv:hep-th/9607207 [hep-th]].



\bibitem{Nakajima:2022sbi}
H.~Nakajima,
\emph{A mathematical definition of Coulomb branches of supersymmetric gauge theories and geometric Satake correspondences for Kac-Moody Lie algebras}, 
[arXiv:2201.08386 [math.RT]].










\bibitem{Teleman:2022oiu}
C.~Teleman,
\emph{Coulomb branches for quaternionic representations},   
[arXiv:2209.01088 [math.AT]].




\bibitem{DP2021}  
D.~Pomerleano,
\emph{Intrinsic mirror symmetry and categorical crepant resolutions},
arXiv:2103.01200 [math.SG].  






\bibitem{Berest}  
Y.~Berest, G.~Khachatryan, A.~Ramadoss, \emph{Derived representation schemes and cyclic homology}, Adv.
Math. 245, 625–689 (2013).




\bibitem{Teleman:2018wac}
C.~Teleman,
\emph{The r\^ole of Coulomb branches in 2D gauge theory}, 
J. Eur. Math. Soc. \textbf{23} (2021) no.11, 3497-3520
doi:10.4171/jems/1071
[arXiv:1801.10124 [math.AG]].   

\bibitem{GIT}
C.~Teleman. \emph{The Quantization Conjecture Revisited.}, Annals of Mathematics 152, no. 1 (2000): 1–43. https://doi.org/10.2307/2661378.






\bibitem{BFM}
R.~Bezrukavnikov, M.~Finkelberg, and I.~Mirkovíc, \emph{Equivariant homology and K-theory of affine Grassmannians and Toda lattices},   
Compos.Math. 141 (2005) 746-768.


\bibitem{FK}
Kirwan, F. C. (1984). Cohomology of Quotients in Symplectic and Algebraic Geometry. (MN-31), Volume 31 (Vol. 104). Princeton University Press. https://doi.org/10.2307/j.ctv10vm2m8

\bibitem{Benini:2010uu}
F.~Benini, Y.~Tachikawa and D.~Xie,
\emph{Mirrors of 3d Sicilian theories},    
JHEP \textbf{09} (2010), 063
doi:10.1007/JHEP09(2010)063
[arXiv:1007.0992 [hep-th]].


\bibitem{Cremonesi:2014vla}
S.~Cremonesi, A.~Hanany, N.~Mekareeya and A.~Zaffaroni,
\emph{Coulomb branch Hilbert series and Three Dimensional Sicilian Theories},
JHEP \textbf{09} (2014), 185
doi:10.1007/JHEP09(2014)185
[arXiv:1403.2384 [hep-th]].


 





\bibitem{Ba}  
Radha Kessar, Markus Linckelmann,
On the Hilbert series of Hochschild cohomology of block algebras,
Journal of Algebra,
Volume 371,
2012,
Pages 457-461,
ISSN 0021-8693,
https://doi.org/10.1016/j.jalgebra.2012.07.020.













  





 




   












  




 


\bibitem{stein}  
\emph{Steinberg slices and group-valued moment maps},
Advances in Mathematics,
402,
108344,
2022,
i0001-8708,
https://doi.org/10.1016/j.aim.2022.108344,
https://www.sciencedirect.com/science/article/pii/S0001870822001608,
Ana Bălibanu.


0\bibitem{Moore:2006dw}
G.~W.~Moore and G.~Segal,
\emph{D-branes and K-theory in 2D topological field theory},   
[arXiv:hep-th/0609042 [hep-th]].  

























\end{thebibliography}
\end{document}